\documentclass [11pt]{article}

\usepackage{amsmath,amssymb,epsfig,cite,color,verbatim}
\usepackage{graphics,float}
\usepackage{overpic} 

\numberwithin{equation}{section}
\newcommand{\TT}{\mathcal{T}}
\newcommand{\zf}{|0_F\rangle}
\newcommand{\zb}{|0_B\rangle}
\newcommand{\EE}{\mathcal{E}}

\title{Light Hadron Masses from a Matrix Model for QCD}
\date{}
\author{Mahul Pandey$^{1}$ and Sachindeo~Vaidya$^2$ \\
${}^1${\small School of Theoretical Physics, Dublin Institute for Advanced Studies, Dublin 4, D04 C932, Ireland}\\
${}^2$ {\small Centre for High Energy Physics,  Indian Institute of Science, Bengaluru, 560012, India}\\
}
\begin{document}
\maketitle
\begin{abstract}

The $SU(3)$ Yang-Mills matrix model coupled to fundamental fermions is an approximation of quantum chromodynamics (QCD) on a 3-sphere of radius $R$. The spectrum of this matrix model Hamiltonian is estimated using standard variational methods, and is analyzed in the strong coupling limit. By employing a matching prescription to determine the dependence of the Yang-Mills coupling constant $g$ on $R$, we relate the asymptotic values of the energy eigenvalues in the $R \rightarrow \infty$ (flat space) limit to the masses of light hadrons. We find that the matrix model estimates the light hadron spectrum fairly accurately, with the light baryon masses falling within $10\%$, and most light meson masses falling within about $30\%$ of their observed values.
\end{abstract}

\section{Introduction}

%

Hadrons constitute almost all of ordinary matter, and the study of their properties is of fundamental importance to test the predictions of low-energy Quantum Chromodynamics (QCD). However, the fact that hadrons lie deep in the nonperturbative regime of the theory poses a considerable challenge for theoretical physicists to predict their properties reliably. Much of our knowledge in this regime comes from numerical simulations in lattice QCD, latest results including \cite{Durr:2008zz,Durr:2008rw,Davies:2003ik, Aoki:2008sm,Bernard:2001av,Aubin:2004wf,Ukita:2007cu,Antonio:2006px,Alexandrou:2008tn,Noaki:2007es,Alexandrou:2017xwd,Alexandrou:2014sha,Bietenholz:2011qq,Borsanyi:2014jba,Jansen:2009hr,Aoki:2011yj} for light hadron spectrum and \cite{Rae:2017xjz,Bali:2012ua,Bali:2015lka,charm3,charm4,charm5,charm6,charm7,charm8} for charmed hadrons. 

A simple and elegant matrix model of Yang-Mills theory, capable of capturing important topological features of the 
full quantum field theory, was obtained in~\cite{Balachandran2014iya, Balachandran2014voa}. The matrix model corresponds to a dimensional reduction of Yang-Mills theory 
on $S^3\times\mathbb{R}$, and is a quantum-mechanical model based on $3\times(N^2-1)$-dimensional real matrices as degrees of freedom. 
This model proves to be quite good in describing the mass spectrum of glueballs \cite{Acharyya:2016fcn} in the low energy regime of pure Yang-Mills theory. 

In this paper, we push the $SU(3)$ matrix model further by coupling fundamental quarks. We provide variational estimates for the masses of light hadrons in this model. Specifically, we give estimates for the masses of the light pseudoscalar and vector mesons and spin-$\frac{1}{2}$ and spin-$\frac{3}{2}$ baryons, which turn out to be surprisingly successful.


We have four unknown parameters in the model, namely, the masses $m_u, m_{d}$ and $m_s$ of the up, down and strange quark, the Yang-Mills coupling $g$ and the radius $R$ of the $3$-sphere, which sets the overall scale in the model. To relate the energy spectrum of the matrix model to physical masses of particles, we need to employ a suitable matching scheme, detailed later in the paper, which involves using the observed masses of four hadrons as inputs. Once the matching scheme has been used to fix the unknown parameters and to set the overall scale, the masses of the other particles can be unambiguously determined from the matrix model.

Although this use of the matrix model in hadron spectroscopy is new, the model itself has appeared earlier in other contexts. For instance, Myers has studied this model to examine the dynamics of $N$ D0-branes in a specific external Ramond-Ramond field \cite{Myers:1999ps}. The energy spectrum of this model, however, has not been investigated thoroughly. We do so here using traditional and well-established variational methods.

The paper is organized as follows. In Section \ref{secMM}, we briefly review the pure gauge matrix model of \cite{Balachandran2014iya, Balachandran2014voa} and describe the inclusion of quarks. In Section \ref{secVar}, we review the variational method for energy eigenvalue determination. In Section \ref{secRenorm}, we set up the mass-matching scheme. Our results are given in Section \ref{secResult}, and a discussion of the results in Section \ref{secDisc}. 

\section{The Matrix Model with Quarks}\label{secMM} 

The matrix model \cite{Balachandran2014iya, Balachandran2014voa} for pure Yang-Mills theory is constructed by isomorphically mapping the spatial $S^3$ to an $SU(2)$ subgroup of the gauge group $SU(N)$, and the pulling back the general left-invariant one-form on $SU(N)$ under this map. The general left-invariant one-form on $SU(N)$ can be expressed as $\Omega = \textrm{Tr}\left( T_a g^{-1} dg\right) M_{ab} T_b$ where $g \in SU(N)$, $T_a$ are the generators in fundamental representation, and $M$ is a $(N^2-1)$-dimensional real matrix. Under the isomorphic mapping, the spatial vector fields  are identified with $i X_i$, where $X_i$ are the left-invariant vector fields in the Lie algebra of $SU(2)$. The pullback of $\Omega$ under this map gives the gauge fields as $A_0 =0, \,A_i = M_{ia} T_a.$ The matrices $M_{ia}$ are $3\times (N^2-1)$ rectangular matrices, and represent our gauge variables. 


The curvature $F_{ij}$ corresponding to $A_i$ is obtained by the pull-back of the curvature associated with $\Omega$ to the spatial $S^3$:
\begin{equation}
F_{ij} = (d \Omega + \Omega \wedge \Omega)(X_i, X_j) = \left(-\frac{1}{R}\epsilon_{ijk}M_{ka}+f_{abc}M_{ib}M_{kc}\right)T_a.
\end{equation}

Then the chromoelectric and chromomagnetic fields read
\begin{eqnarray}
E_{ia} &\equiv& F_{0ia}=\dot{M_{ia}}, \\
B_{ia} &\equiv& \frac{1}{2}\epsilon_{ijk}F_{jka}=-\frac{1}{R}M_{ia}+\frac{1}{2}\epsilon_{ijk}f_{abc}M_{jb}M_{kc}.
\end{eqnarray}

We ignore the numerical factor in the volume of $S^3$ and take it to be $R^3$. The matrix model Lagrangian is
\begin{equation}
L_{YM}= -\frac{R^3}{4g^2}F_{\mu\nu}^aF^{a\mu\nu}=\frac{R^3}{2g^2} \left( E_i^a E_i^a -B_i^a B_i^a \right).
\end{equation}

In the temporal gauge $A_0=0$, the Lagrangian is expanded to obtain quadratic, cubic and quartic interactions in the potential:
\begin{equation}
L_{YM}=\frac{R^3}{2g^2}\left(\dot{M}_{ia}\dot{M}_{ia}-\frac{1}{R^2} M_{ia}M_{ia}+\frac{1}{R}\epsilon_{ijk}f_{abc}M_{ia}M_{jb}M_{kc}-\frac{1}{2}f_{abc}f_{ade}M_{ib}M_{jc}M_{id}M_{je}\right).
\label{LMM}
\end{equation}

To the matrix model Lagrangian (\ref{LMM}), we add minimally coupled massive 3-flavor quarks $Q$ in the Dirac representation, transforming in the fundamental representation of color $SU(3)$, and flavor $SU(3)$. The spinor $Q \equiv (Q_{f \alpha A})$ carries indices $f,\alpha$ and $A$ denoting flavour, spin and colour indices respectively.
Under gauge transformations $u\in SU(3)$, spatial rotations $R \in SO(3)$ and flavour rotations $v \in SU(3)$, $Q_{f \alpha A}$ transforms 
as
\begin{eqnarray}
Q_{f \alpha A} &\rightarrow& u_{AB} Q_{f \alpha B}, \quad A, B = 1,2,3\\
Q_{f \alpha A} &\rightarrow& D^{\frac{1}{2}}(R)_{\alpha \beta} Q_{f \beta A}, \quad \alpha, \beta =1,2\\
Q_{f \alpha A} &\rightarrow& v_{fg} Q_{g \alpha A}, \quad f,g=1,2,3
\end{eqnarray}
where $D^{\frac{1}{2}}(R)_{\alpha \beta}$ is the spin-$\frac{1}{2}$ representation of $R$.

A Dirac fermion $Q$ is made up of a left Weyl fermion $b$ and a right Weyl fermion $c^\dagger$:
\begin{equation}
Q_{f \alpha A} = \left(\begin{array}{c}
			b_{f \alpha A} \\
			-i (\sigma_2)_{\alpha \beta} c^\dagger_{f \beta A} \end{array} \right) \equiv 
			\left(\begin{array}{c}
			b_{f \alpha A} \\
			d^\dagger_{f \alpha A} \end{array} \right).
\end{equation}
This implies that $b$ transforms in the fundamental and $d$ in the anti-fundamental representations of color and flavor, and both have the same transformation property under rotations:
\begin{eqnarray}
b_{f \alpha A} &\rightarrow& u_{AB} b_{f \alpha B}, \quad A, B = 1,2,3\\
b_{f \alpha A} &\rightarrow& D^{\frac{1}{2}}(R)_{\alpha \beta} b_{f \beta A}, \quad \alpha, \beta =1,2\\
b_{f \alpha A} &\rightarrow& v_{fg} b_{g \alpha A}, \quad f,g=1,2,3
\end{eqnarray}
and
\begin{eqnarray}
d_{f \alpha A} &\rightarrow& u^*_{AB} d_{f \alpha B}, \quad A, B = 1,2,3\\
d_{f \alpha A} &\rightarrow& D^{\frac{1}{2}}(R)_{\alpha \beta} d_{f \beta A}, \quad \alpha, \beta =1,2\\
d_{f \alpha A} &\rightarrow& v^*_{fg} d_{g \alpha A}, \quad f,g=1,2,3
\end{eqnarray}

The Lagrangian for minimally coupled massive quarks on a sphere is given by \cite{Senferm}
\begin{equation}
L=-\frac{R^3}{4g^2}F_{\mu\nu}^aF^{a\mu\nu}
+R^3\left(i\bar{Q}_{f} \gamma^\mu (\mathcal{D}_\mu Q_f)
-m_f\bar{Q}_f Q_f - \frac{3}{2R}\bar{Q}_f\gamma^5\gamma^0 Q_f \right)
\label{LYMD}
\end{equation}
where $\bar{Q}=Q^\dagger\gamma^0$. The last term in eq. (\ref{LYMD}) comes from the non-zero curvature of the spatial $S^3$.

Both the gauge fields $M_{ia}$ and the quarks $b$ and $d$ depend only on time. $M$ has dimensions of inverse length. To express the Hamiltonian in terms of dimensionless quantities and for computational ease, it is convenient to rescale $M\rightarrow \frac{gM}{R}$. Also, to make the fermionic variables dimensionless, we rescale $b\rightarrow R^{-\frac{3}{2}} b,d\rightarrow R^{-\frac{3}{2}} d$, and express the quark masses in units of $R^{-1}$, $m_f \equiv \frac{\mu_f}{R}$. The Hamiltonian then works out to be
\begin{equation}
H=\frac{1}{R}(H_{YM}+H_m+H_c+H_{int})
\label{hamil}
\end{equation}
where
\begin{equation}
H_{YM}=\frac{1}{2}\Pi_{ia}\Pi_{ia}+\frac{1}{2} M_{ia}M_{ia}-\frac{g}{2}\epsilon_{ijk}f_{abc}M_{ia}M_{jb}M_{kc}+\frac{g^2}{4}f_{abc}f_{ade}M_{ib}M_{jc}M_{id}M_{je},
\label{hym}
\end{equation}
\begin{equation}
H_m=\sum_{f} \mu_f(b^{\dagger}_{f\alpha A}b_{f\alpha A}-d_{f \alpha A}d^\dagger_{f\alpha A}),
\label{hm}
\end{equation}
\begin{equation}
H_c=\frac{3}{2}(b^\dagger_{f\alpha A}d^\dagger_{f\alpha A}+d_{f\alpha A}b_{f\alpha A}),
\label{hc}
\end{equation}
\begin{equation}
H_{int}=-\frac{g}{2} M_{ia}(b^\dagger_{f\alpha A} \sigma^i_{\alpha\beta}\lambda^a_{AB}d^\dagger_{f\beta B}+d_{f\alpha A} \sigma^i_{\alpha\beta}\lambda^a_{AB}b_{f\beta B}).
\label{hint}
\end{equation}
Here $\lambda^a$ are the Gell-Mann matrices.

The Gauss' law constraint is
\begin{equation}
G_a=f_{abc}\Pi_{ib}M_{ic}-\frac{1}{2}(b_f^\dagger \lambda_a b_f- d_f \lambda_a d_f^\dagger),
\end{equation}
where the suppressed indices are understood as being summed over.

To quantize the system, we impose the canonical commutation (and anti-commutation) relations
\begin{equation}
\left[M_{ia},\Pi_{jb}\right]=i\delta_{ij}\delta_{ab}, \quad\quad\{ b_{f\alpha A},b^\dagger_{g\beta B} \}=\delta_{\alpha\beta}\delta_{AB}\delta_{fg},\quad\quad\{ d_{f\alpha A},d^\dagger_{g\beta B} \}=\delta_{\alpha\beta}\delta_{AB}\delta_{fg}
\end{equation}
and demand that all physical states $|\Psi\rangle_{phys}$ be annihilated by the Gauss law:
\begin{equation}
G_a |\Psi\rangle_{phys} =0.
\end{equation}

Total spin, given by
\begin{equation}
J_i=\epsilon_{ijk}M_{ja}\Pi_{ka}+\frac{1}{2}(b_f^\dagger \sigma_i b_f- d_f \sigma_i d_f^\dagger),
\end{equation}
commutes with $H$ and so its eigenstates can be assigned a spin unambiguously.

In the fundamental representation, the generators of flavour symmetry are given by $\mathcal{T}_F\equiv\frac{1}{2}\lambda_F, F=1,...,8$, $\lambda_F$ being the Gell-Mann matrices. In terms of these generators, we can define the isospin operator $I_3$ and the hypercharge $Y$ as
\begin{align}
& I_3=Q^\dagger\mathcal{T}_3 Q = b^\dagger \mathcal{T}_3 b+d \mathcal{T}_3 d^\dagger,\\
& Y=\frac{2}{\sqrt{3}}Q^\dagger\mathcal{T}_8 Q =\frac{2}{\sqrt{3}} (b^\dagger \mathcal{T}_8 b+d \mathcal{T}_8 d^\dagger).
\end{align}
These operators also commute with $H$ and so the eigenstates can be assigned isospin and hypercharge. 

We will also need the fermion number operator
\begin{equation}
N_Q=b^\dagger b+d^\dagger d.
\end{equation}
Eigenstates of $N_Q$ have a definite number of quarks $+$ antiquarks.

To estimate light hadron masses, we choose a set of variational states $\{|\Psi_n\rangle\}$ and organize them according to their spin and flavour quantum numbers. Then, we compute $\widetilde{H}_{mn}\equiv \langle \Psi_m|H|\Psi_n\rangle$ and the Gram matrix $S_{mn}=\langle\Psi_m|\Psi_n\rangle$, defined in section \ref{secVar}, and evaluate the generalized eigenvalues and eigenvectors of $\widetilde{H}$ with respect to $S$. The eigenvalues have functional dependence on the bare parameters of the theory, namely, the quark masses $\mu_f$ and the coupling constant $g$, as well as the infra-red cut-off $R$. To obtain well-defined values for light hadron masses, we need to prescribe a suitable renormalization scheme to determine the flow of the bare parameters with the scale $R$. A brief description of this renormalization procedure is given in the section \ref{secRenorm}. 

\section{Variational Calculation}\label{secVar}

To estimate the eigenvalue spectrum of the matrix model Hamiltonian, we use Rayleigh-Ritz variational method. We briefly summarize the steps for the variational calculation in the following paragraphs.

We first construct a trial set of variational states $\{|\psi_i\rangle\}$. Then we consider the following variational ansatz: 
\begin{equation}
|\chi\rangle = \sum_{i} c_i |\psi_i\rangle, \quad c_i \in \mathbb{C}. 
\end{equation}
To estimate the spectrum, consider the functional 
\begin{equation}
\mathcal{K}= \langle \chi |H | \chi \rangle - \lambda (\langle\chi|\chi \rangle -1)
\end{equation}
where $\lambda$ is a Lagrange multiplier for the constraint that the state $|\chi\rangle$ is normalized to 1. 

Minimizing this functional wrt $c_i^{\ast}$ leads to the generalized eigenvalue equation,
\begin{equation}
\sum_{j}\widetilde{H}_{ij}c_j =\lambda \sum_{j}  \widetilde{S}_{ij} c_j,
\label{evaleqn}
\end{equation}
where $\widetilde{H}_{ij}\equiv \langle \psi_i |H|\psi_j\rangle$ and $S_{ij} \equiv \langle \psi_i |\psi_j\rangle.$

Minimizing wrt $\lambda$ yields
\begin{equation}
\sum_{ij} c_i^{*}S_{ij}c_j=0.
\label{seqn}
\end{equation}

Equations (\ref{evaleqn}) and (\ref{seqn}) together give
\begin{align}
\lambda &=\frac{\sum_{jk}c_k^{*}\widetilde{H}_{kj}c_j}{\sum_{jk}c_k^{*}S_{kj}c_j}=\frac{\langle \chi|H|\chi\rangle}{\langle \chi|\chi\rangle}.
\end{align}
Thus, the lowest eigenvalue of the matrix $\widetilde{H}$ wrt $S$ gives the ground state energy and the higher eigenvalues give the energies of the excited states. Furthermore, the variational states can be organized via their quantum numbers under observables that commute with $H$, for example total spin, isospin and hypercharge. $\widetilde{H}$ then assumes a block-diagonal form, and each block can be treated separately.

A typical variational state in our ansatz is a combination of a fermionic and a bosonic wavefunction. The fermionic part of the wavefunction can be created by successive operation of the fermion and antifermion creation operators $b^\dagger$ and $d^\dagger$ respectively on the fermion Fock vacuum, denoted by $\zf$. As the bosonic part of the wavefunction, we take eigenfunctions of the $24$-dimensional bosonic oscillator, given by
\begin{equation}
H_{osc}\equiv\frac{1}{2}\left(\Pi_{ia}\Pi_{ia}+M_{ia}M_{ia} \right).
\label{hosc}
\end{equation}

To construct our variational ansatz, we take combinations of different bosonic and fermionic wavefunctions such that the combinationa as a whole transforms as a colour singlet. 
In this article, we consider fermionic states with up to five fermions, and bosonic states with up to four bosonic oscillators.  They are combined in different ways to give $449$ variational states in the ansatz for scalar mesons, $1017$ states for vector mesons, $449$ states for the baryon octet and $604$ states for the baryon decuplet.

The procedure for constructing the variational states is outlined briefly in Appendix A, where we also list all linearly independent fermionic and bosonic wavefunctions in our ansatz.

\section{Light Hadrons from the Spectrum of H}\label{secRenorm}

The energy eigenvalues of $H$ in general have functional dependence on four parameters in our theory, namely, the quark masses $\mu_f, f=1,...,3$ and the coupling constant $g$. As an approximation, we take the bare masses of the up and down quarks to be the same. Thus, we have two mass parameters, $\mu\equiv\mu_{u/d}$ and $\mu_s$, and the coupling constant $g$. We also have a length scale in our theory, which is the infrared cut-off $R$. The bare parameters $\mu$, $\mu_s$ and $g$ have no physical meaning. To obtain meaningful results in the flat-space, or $R\rightarrow\infty$ limit, the bare parameters are chosen to depend on $R$ in such a way that the energy eigenvalues asymptote to fixed numbers at large $R$.

The model takes into account only the zero mode sector of the full quantum field theory. The (renormalized) zero-point energy of the higher momentum modes of the full quantum field theory can at most change the spectrum of our Hamiltonian by an overall additive constant $C(R)$. Energy differences, of course, will not depend on $C(R)$.

Energy eigenvalues of (\ref{hamil}), taking into account the shift $C(R)$, have the functional form
\begin{equation}
\mathcal{E}_n^k(\mu,\mu_s,g,R)\equiv \frac{f_n^k(\mu,\mu_s,g)}{R}+C(R)
\end{equation}
where $k$ denotes the quantum numbers (spin, isospin etc.) assigned to the corresponding eigenstates, and $n$ labels all the eigenstates with the same quantum numbers. 

Let us look at the eigenvalues corresponding to four particles labeled by $(k_i,n_i),i=1,...,4$. We can define two ratios of the energy differences, which has functional dependence only on $\mu,\mu_s,g$:
\begin{equation}
\mathcal{R}_{123}(\mu,\mu_s,g)\equiv\frac{f_{n_1}^{k_1}-f_{n_2}^{k_2}}{f_{n_1}^{k_1}-f_{n_3}^{k_3}},
\end{equation}
\begin{equation}
\mathcal{R}_{423}(\mu,\mu_s,g)\equiv\frac{f_{n_4}^{k_4}-f_{n_2}^{k_2}}{f_{n_1}^{k_1}-f_{n_3}^{k_3}}.
\end{equation}

We now demand that the two ratios be fixed to their observed values for all $g$. In other words, we look for solutions to the set of equations,
\begin{equation}
\mathcal{R}_{123}(\mu,\mu_s,g)=R^{obs}_{123},\quad \mathcal{R}_{423}(\mu,\mu_s,g)=R^{obs}_{423},
\label{scaling}
\end{equation}
where $R^{obs}$ denotes the observed values of the two ratios. Solving $(\ref{scaling})$ gives us two functions $\mu(g)$ and $\mu_s(g)$. With this choice of $\mu(g)$ and $\mu_s(g)$, the ratio of the energy differences between any two pairs of particles can be computed. In particular, let us define
\begin{equation}
\mathcal{R}_x(g)\equiv \frac{f_{n_x}^{k_x}-f_{n_2}^{k_2}}{f_{n_1}^{k_1}-f_{n_3}^{k_3}}
\end{equation}
where $x\neq1,2,3,4$. We observe that for each $x$, $\mathcal{R}_x(g)$ approaches a well-defined asymptotic value for large $g$. These ratios are predictions of our model and listed in Table \ref{tablerat}.

Let us now see how to determine the mass differences themselves. To this end, let us look at a particular energy difference, say between particles $1$ and $2$, and substitute the functions $\mu(g)$ and $\mu_s(g)$ in its expression:
\begin{equation}
\Delta \mathcal{E}_{12}(g,R)\equiv \mathcal{E}_{n_1}^{k_1}-\mathcal{E}_{n_2}^{k_2}= \frac{f_{n_1}^{k_1}(g,\mu(g),\mu_s(g) )-f_{n_2}^{k_2}(g,\mu(g),\mu_s(g) )}{R}
\end{equation}

We again demand that  for all $R$, $\Delta \mathcal{E}_{12}$ be fixed to $\Delta M_{12}^{obs}$, the observed value of the mass difference of particles 1 and 2. $\Delta M_{12}^{obs}$ is given in terms of a physical unit $l$ (like MeV$^{-1}$). Then, in terms of $x\equiv \frac{R}{l}$, the energy difference is
\begin{equation}
\Delta \mathcal{E}_{12}(g,x)=\left(\frac{f_{n_1}^{k_1}(g)-f_{n_2}^{k_2}(g)}{x}\right)\frac{1}{l}.
\end{equation}
The equation
\begin{equation}
\Delta \mathcal{E}_{12}(g,x)=\Delta M_{12}^{obs}.
\label{scaling2}
\end{equation}
implicitly determines the function $g(x)$, which in turn determines $\mu(x)=\mu(g(x))$ and $\mu_s(x)=\mu_s(g(x))$. We will refer to (\ref{scaling}) and (\ref{scaling2}) as mass-matching conditions.

 In practice, it is easier to determine $x$ as a function of $g$: inverting (\ref{scaling2}), we obtain
\begin{equation}
x(g)= \left(\frac{f_{n_1}^{k_1}(g)-f_{n_2}^{k_2}(g)}{\Delta E_{12}^{obs}}\right)\frac{1}{l}.
\end{equation}
This enables us to express the mass difference of a particle $x$ and particle $2$ in terms of the corresponding mass difference ratio $\mathcal{R}_x$:
\begin{equation}
\Delta \mathcal{E}_{x,2}(g)\equiv\mathcal{E}_{n_x}^{k_x}(g)-\mathcal{E}_{n_2}^{k_2}(g)= \mathcal{R}_x(g)\Delta E_{12}^{obs}.
\end{equation}

Finally, to determine the actual masses of our particles, we set the asymptotic value of $\mathcal{E}_{n_2}^{k_2}$ to the observed mass of that particle from experimental data. This amounts to fixing the value of the additive constant $C$, which is the only parameter left in the theory to adjust.

Thus, by choosing the observed masses of four particles as inputs, we have determined the four functions, namely, $\mu(x),\mu_s(x),g(x)$ and $C(x)$. This enables us to predict the absolute masses of the other particles unambiguously.

\section{Results}\label{secResult}

\subsection{Identification of the Particles}
Once we have determined the variational Hamiltonian $\widetilde{H}$ of (\ref{evaleqn}), we proceed to compute its generalized eigenvalues and eigenvectors, and identify the quantum numbers associated with them. Parity is not a good symmetry of the Hamiltonian due to the cubic term in $H_{YM}$ and the fermion curvature term $H_c$. We do note that the expectation value of the parity operator in any eigenstate does asymptote to $\pm 1$ in the large $g$ limit, but this value is sensitive to the variational ansatz. Thus we cannot assign parity unambiguously in the current level of approximation.

\begin{figure}[!hbtp]
\centering
\includegraphics[scale=0.6]{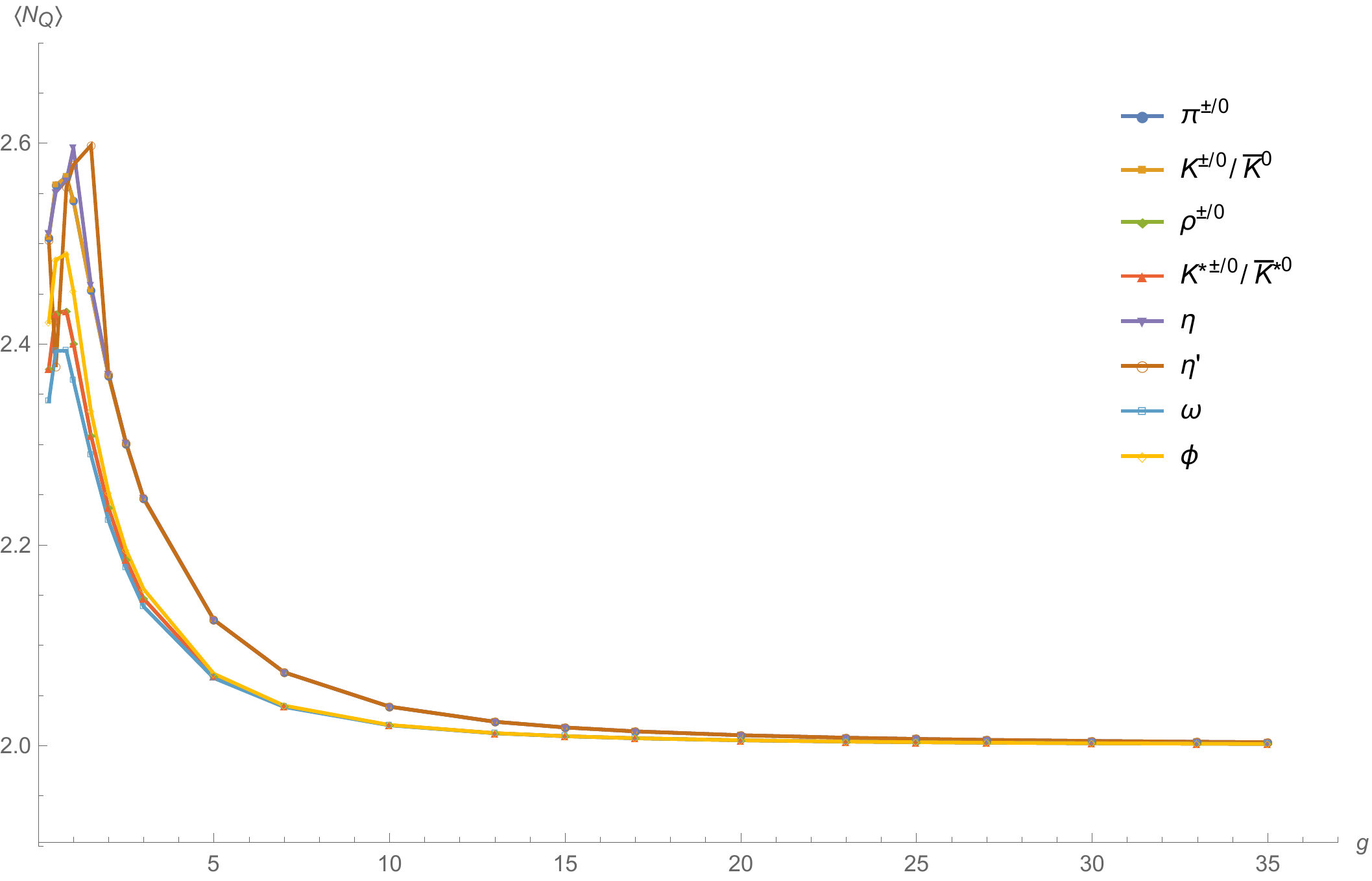}
\caption{A plot of $\langle N_Q\rangle$ vs $g$ for the mesons}
\label{fignq1}
\end{figure}

\begin{figure}[!hbtp]
\centering
\includegraphics[scale=0.6]{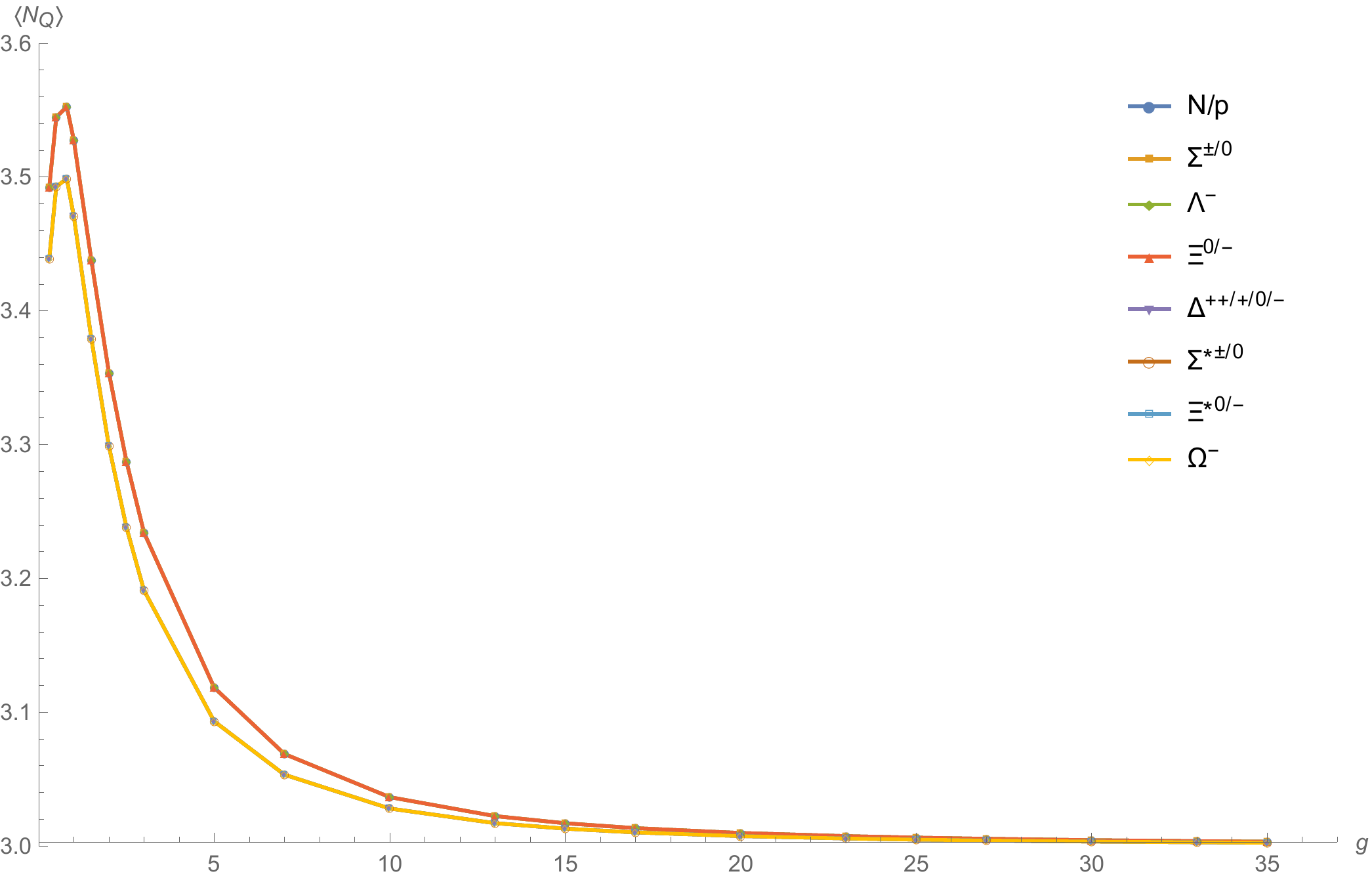}
\caption{A plot of $\langle N_Q\rangle$ vs $g$ for the baryons}
\label{fignq2}
\end{figure}

We can, however, make two important qualitative predictions from our model:
\begin{enumerate}

\item The Hamiltonian is symmetric under charge conjugation, which takes $M_{ia}T_a\rightarrow -M_{ia}T_a^*$ and $Q\rightarrow i\gamma^2Q^*$. Therefore particles and their antiparticles have the same energy. Similarly, the expectation values of various other quantum numbers, namely, spin, isospin and hypercharge are well-defined for all $g$, because the corresponding generators commute with $H$. This enables us to assign these quantum numbers to the states in our multiplets. As a first qualitative test of our model, we find that the quantum numbers assigned to the lowest flavour multiplets, in both the baryonic and mesonic sector, follow the same mass hierarchy, and \textit{exactly} match with the corresponding particles observed in nature.

\item Any variational eigenstate is a linear combination of states with different number of quarks and antiquarks. Nevertheless, the expectation value of the fermion number operator $N_Q$ in such an eigenstate is well-defined. We find that in the large $g$ limit, the lowest flavour multiplet in the mesonic sector has $\langle N_Q\rangle=2.00042$ and in the baryonic sector has $\langle N_Q\rangle=3.00040$. We thus identify them with the lightest mesons and baryons respectively. We illustrate this for the pseudoscalar and vector meson octets in Fig. \ref{fignq1}, and baryon octet and decuplet in Fig. \ref{fignq2}.

\end{enumerate} 

\subsection{Mass-Matching, Predictions and Error Analysis}

Following the argument outlined in section \ref{secRenorm}, we choose the observed masses of the pion, the $\rho$-meson, the $\Lambda$ baryon and the $\Delta$ baryon as inputs and solve (\ref{scaling}) and (\ref{scaling2}). We take their experimental masses from the particle data booklet \cite{pdb2018}: $m_\pi=138.04$ MeV, $m_\rho=775.16$ MeV, $m_\Lambda=1115.16$ MeV and $m_\Delta=1232.00$ MeV. Because we have taken the masses of the $u$ and $d$ quarks to be the same, isospin symmetry is exact, and particles in the same isospin multiplet have equal masses. Thus the experimental numbers quoted are the average value of the masses of all particles in the same isospin multiplet.

%
%

We first determine the functions $\mu(g)$ and $\mu_s(g)$ by fixing two mass-difference ratios to their experimental values:

\begin{equation}
\mathcal{R}_{\rho\pi\Lambda}(\mu,\mu_s,g)\equiv\frac{f_{\Lambda}-f_{\pi}}{f_{\rho}-f_{\pi}}=1.534,
\label{r1}
\end{equation}
\begin{equation}
\mathcal{R}_{\rho\pi\Delta}(\mu,\mu_s,g)\equiv\frac{f_{\Delta}-f_{\pi}}{f_{\rho}-f_{\pi}}=1.717
\label{r2}
\end{equation}

\begin{figure}[hbtp]
\centering
\includegraphics[scale=0.8]{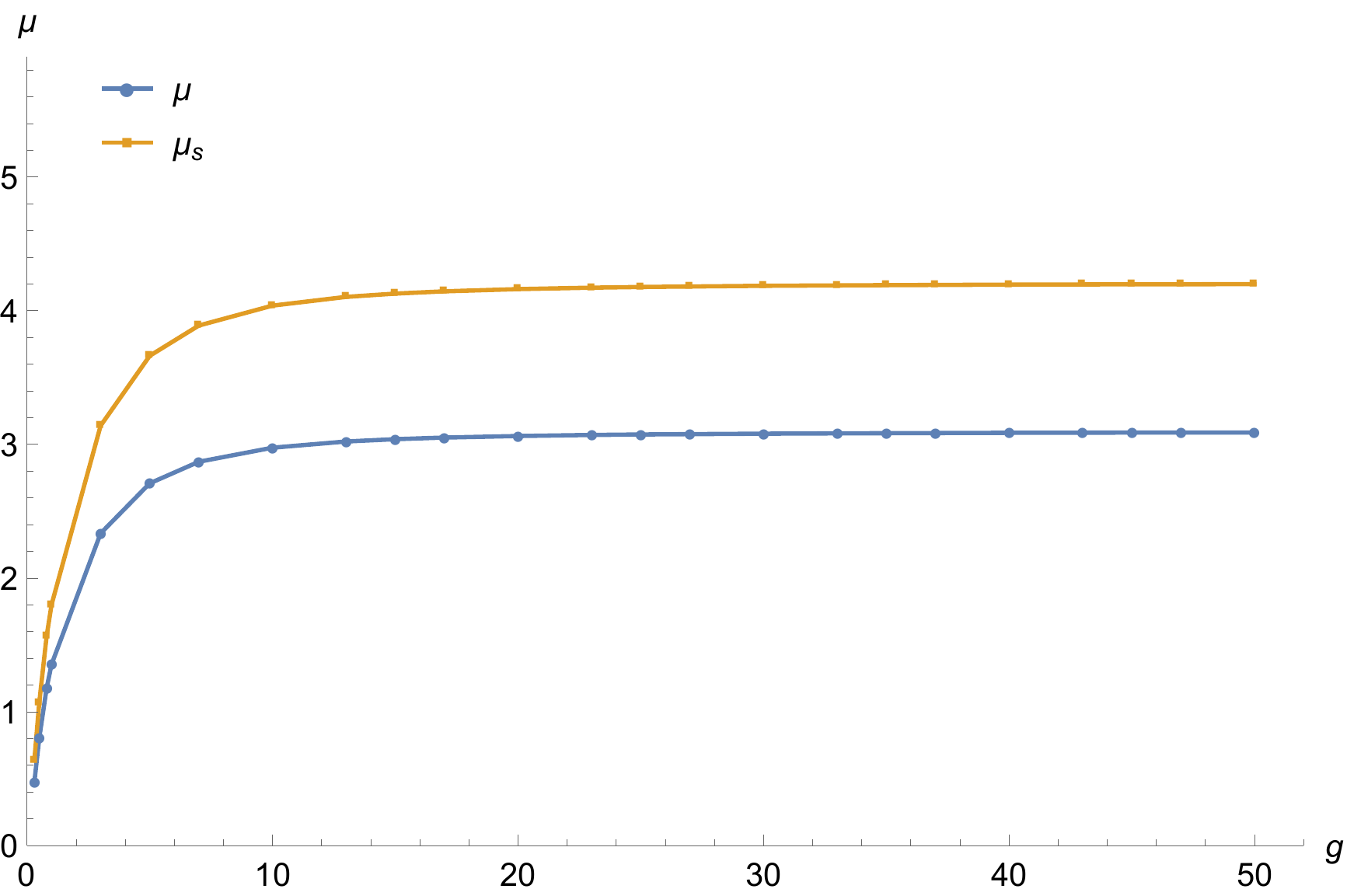}
\caption{A plot of $\mu$ and $\mu_s$ vs $g$}
\label{figmu}
\end{figure}

Since (\ref{r1}) and (\ref{r2}) are non-linear equations involving eigenvalues of very large matrices, it is not easy to obtain analytic solutions. However it is fairly easy to solve them numerically for a discrete set of values of $g$. As shown in Fig. \ref{figmu}, the functions $\mu(g)$ and $\mu_s(g)$ asymptote to definite values for large $g$.

These asymptotic values give the constituent quark masses in our model, in units of inverse $R$. The ratio of the constituent quark masses is independent of the unit chosen, and is found to be
\begin{equation}
\frac{\mu_s}{\mu}=1.26.
\end{equation}

With this choice of scaling functions $\mu(g)$ and $\mu_s(g)$, the other mass-difference ratios also asymptote to constant values. We demonstrate this graphically for a few of the particles in Fig. \ref{figrat} below.

\begin{figure}[hbtp]
\centering
\includegraphics[scale=0.8]{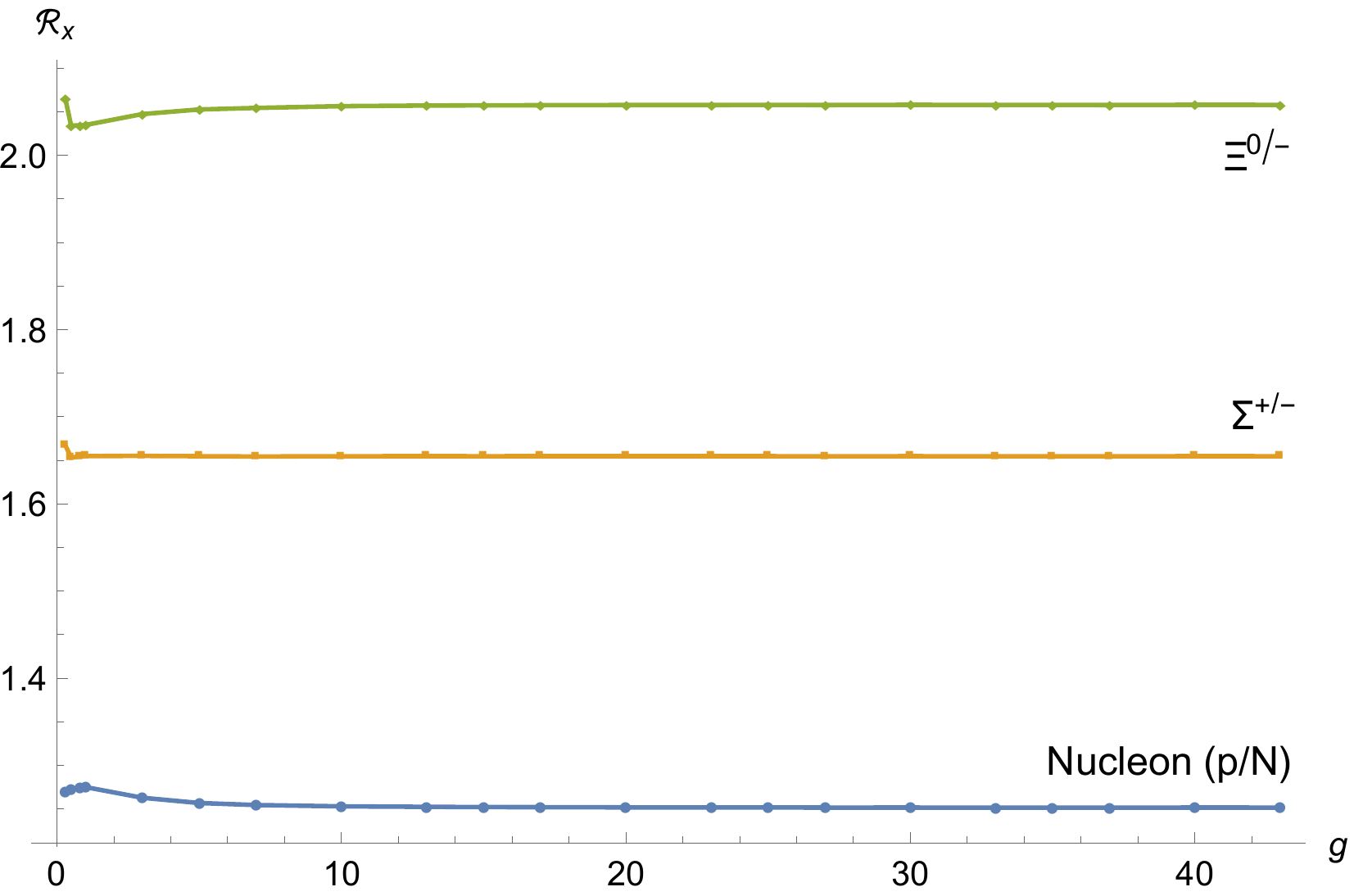}
\caption{Plot of $\mathcal{R}_x$ vs $g$ for some particles $x$. We show the asymptotic behaviour of $\mathcal{R}_x$ for three of the baryons, namely, the nucleon, the $\Lambda^0$ and the $\Xi$ baryon.}
\label{figrat}
\end{figure}

To determine the actual masses, we use $m_\pi$=138.04 MeV and $m_\rho$=775.16 MeV, and predict the masses of the other hadrons.

Before we report our predictions, we would like to say a few words about the choice of input masses and error analysis. Different sets of input particles lead to slightly different predictions for the masses of the other particles. For each input set, we can compute the chi-squared $(\chi^2)$ value of our predictions, with $\chi^2$ defined as
\begin{equation}
\chi^2\equiv \sum_i\frac{(M^i_{mm}-M^i_{obs})^2}{M^i_{obs}}.
\end{equation}
Here, $M^i_{mm}$ and $M^i_{obs}$ are the predicted and observed masses of the $i^{th}$ particle respectively. In Table \ref{T1} below, we provide a list of typical $\chi^2$ values for various choices of inputs. This justifies our choice of $\pi,\rho,\Lambda,\Delta$ as inputs because it minimizes $\chi^2$. 
\begin{table}[!h]
\centering
\begin{tabular}{|c|c|}
\hline
Inputs & $\chi^2$ (MeV)\\ 
\hline 
$\pi,\rho,\Lambda,\Delta$ &  $95.38$ \\ 
\hline 
$\pi,\rho,\Sigma^*,\Delta$ &  $104.31$ \\ 
\hline 
$\pi,\rho,\Xi,\Delta$ & $107.29$ \\ 
\hline
$\pi,\rho,\Xi^*,\Delta$ & $109.49$ \\ 
\hline
$\pi,\rho,\Omega,\Delta$ & $126.09$ \\ 
\hline
$\pi,\rho,\Lambda,\Xi$ & $132.17$ \\ 
\hline 
$\rho,K^*,\Lambda,\Delta$ & $191.13$ \\ 
\hline

\end{tabular} 
\caption{A list of input sets giving lowest few values of $\chi^2$}
\label{T1}
\end{table}

In the above discussion, we have set the values of our input masses to their exact quoted observed values. However, in principle, there are two important sources of mismatch of our model with nature. One source of mismatch is of course the truncation of our variational ansatz at the level of four-boson and five-fermion states. The other source of mismatch lies in the fact that the matrix model is an approximation of the full QCD. These two sources of mismatch effectively induce error bars in the input masses, which in turn propagate and give error bars for our predictions. 


Although quantitative error analysis of variational estimates is notoriously difficult, we can make a rough estimate of the input errors by varying our input set. We do this in the following way: we choose a larger set of input particles, namely, $\pi,\rho,\Lambda,\Delta, \Sigma^*$ and $\Xi$. Out of this set, we can make 15 independent choices of $4$ inputs, each of which leads to a slightly different prediction for the mass of the rest of the particles in this set. This enables us to put error bars in the masses of our original choice of input particles, which we summarize in Table \ref{T2}.

\begin{table}[!h]
\centering
\begin{tabular}{|c|c|}
\hline
Input Particle & Mass with Error Bar (MeV)\\ 
\hline 
$\pi^{\pm/0}$ &  $138.04\pm37.88$ \\ 
\hline 
$\rho^{\pm/0}$ &  $775.16\pm51.53$ \\ 
\hline 
$\Lambda^-$ & $1115.16\pm21.52$ \\ 
\hline
$\Delta^{++/+/0/-}$ & $1232.00\pm28.18$ \\ 
\hline
\end{tabular} 
\caption{A list of input masses with error bars}
\label{T2}
\end{table}
 With these error bars, the ratio of constituent quark masses is obtained as

\begin{equation}
\frac{\mu_s}{\mu}=1.260\pm0.047.
\end{equation}

\subsection{Results}

\begin{figure}[hbtp]
\begin{center}
\begin{overpic}[width=1\hsize]{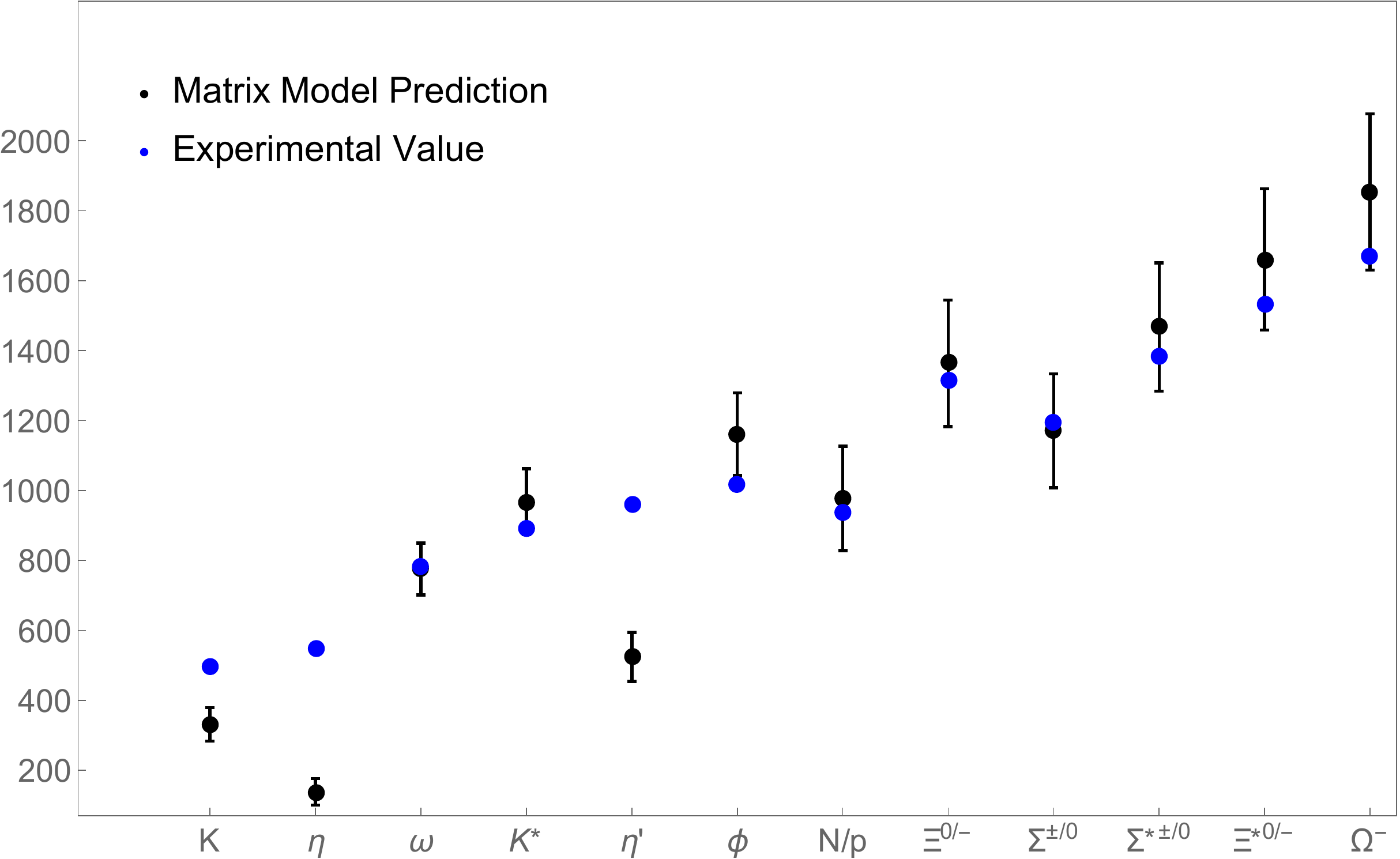}
\put(-4,15){\rotatebox{90}{Masses (MeV)}}
\put(40,-3){\rotatebox{0}{Hadrons}}
\end{overpic}
\end{center}
\caption{{Matrix Model estimates of the light hadron masses compared with observed values.}}
\label{figc}
\end{figure}

Our results are given in Tables \ref{tablerat}, \ref{tablemes} and \ref{tablebar}. In Table \ref{tablerat}, we report the asymptotic values of the mass diference ratios $\mathcal{R}_x$ for all the light mesons
and baryons (except the ones used as inputs) from the matrix model, as well as their observed values. In Tables \ref{tablemes} and \ref{tablebar}, we present the matrix model predictions for the light meson and baryon masses from our model and compare these with the observed values taken from the particle data booklet \cite{pdb2018}. We also give the percentage error by which our results (central values) differ from the observed values. Fig. \ref{figc} gives a summary of our results.

\begin{table}[!h]
\begin{tabular}{|c|c|c|}
\hline 
Particle (x) & $\mathcal{R}_x=\frac{\mathcal{E}_x-\mathcal{E}_\pi}{\mathcal{E}_\rho-\mathcal{E}_\pi}$ & Observed value of $R_x=\frac{m_x-m_\pi}{m_\rho-m_\pi}$ \\ 
\hline
$K^{\pm}/K^0/\bar{K}^0$& 0.303$\pm$0.035 & 0.560\\ 
\hline 
$K^{*\pm}/K^{*0}/\bar{K}^{*0}$& 1.303$\pm$0.035 & 1.185 \\ 
\hline 
$\eta$ & 0.001$\pm$0 & 0.643 \\ 
\hline 
$\eta'$& 0.606$\pm$0.070 & 1.267 \\ 
\hline 
$\omega$& 1.001$\pm$0 & 1.011\\ 
\hline 
$\phi$ & 1.606$\pm$0.070 & 1.385\\ 
\hline 
$p/N$& 1.318$\pm$0.184 & 1.256 \\
\hline 
$\Xi^{0/-}$& 1.924$\pm$0.200 & 1.846 \\ 
\hline 
$\Sigma^0$ & 1.621$\pm$0.188 & 1.655 \\ 
\hline 
$\Sigma^{*(\pm/0)}$ & 2.087$\pm$0.188 & 2.190 \\ 
\hline 
$\Xi^{*0/-}$ & 2.390$\pm$0.199 & 1.958 \\ 
\hline 
$\Omega^-$ & 2.693$\pm$0.215 & 2.408 \\ 
\hline
\end{tabular} 
\caption{Mass-difference ratios with error bars for different mesons and baryons, and their observed values}
\label{tablerat}
\end{table}

\begin{table}[!hbtp]
\centering
\begin{tabular}{|c|c|c|c|c|c|}
\hline
Particle & Spin & Isospin & Matrix Model Mass (MeV) & Observed Mass (MeV) & $\%$ Difference\\ 
\hline
$K^{\pm}/K^0/\bar{K}^0$& 0 & 1/2 & 331.09$\pm$48.06 & 495.65 & -33.2 $\%$ \\ 
\hline 
 $\eta$ & 0 & 0 & 138.03$\pm$37.88 & 547.86 & -74.8 $\%$ \\ 
\hline 
$\eta'$& 0 & 0 & 524.14$\pm$70.23 & 957.78 & -45.3 $\%$ \\ 
\hline 
$K^{*\pm}/K^{*0}/\bar{K}^{*0}$& 1 & 1/2 & 968.21$\pm$94.22 & 893.65 & +8.3 $\%$ \\ 
\hline 
$\omega$& 1 & 0 & 775.19$\pm$74.33 & 782.65 & -0.9 $\%$ \\ 
\hline 
$\phi$ & 1 & 0& 1161.24$\pm$118.23 & 1019.46 & +13.9 $\%$ \\
\hline 
\end{tabular} 
\caption{Comparison of the meson masses in the matrix model with data}
\label{tablemes}
\end{table}

\begin{table}[!hbtp]
\centering
\begin{tabular}{|c|c|c|c|c|c|}
\hline
Particle & Spin & Isospin & Matrix Model Mass (MeV) & Observed Mass (MeV) & $\%$ Difference\\ 
\hline
$p/N$& 1/2 & 1/2 & 977.74$\pm$149.24 & 938.91 & +4.1 $\%$ \\ 
\hline 
$\Xi^{0/-}$& 1/2 & 1/2 & 1363.85$\pm$180.69 & 1314.86 & +3.7 $\%$ \\ 
\hline 
$\Sigma^{(\pm/0)}$ & 1/2 & 0 & 1170.79$\pm$163.05 & 1193.15 & -1.9 $\%$ \\ 
\hline 
$\Sigma^{*(\pm/0)}$ & 3/2 & 1 & 1467.84$\pm$183.46 & 1384.6 & +6.0 $\%$ \\ 
\hline 
$\Xi^{*0/-}$ & 3/2 & 1/2 & 1660.89$\pm$202.18 & 1533.4 & +8.3 $\%$ \\ 
\hline 
$\Omega^-$ & 3/2 & 0 & 1853.94$\pm$223.25 & 1672.45 & +10.8 $\%$ \\ 
\hline 
\end{tabular} 
\caption{Comparison of the baryon masses in the matrix model with data}
\label{tablebar}
\end{table}

\newpage

\section{Discussion}\label{secDisc}
We observe that the matrix model predicts light hadron masses with fair accuracy, with most masses lying within 20$\%$ of their observed values. The predictions for baryon masses are especially good, with surprisingly small errors.


We do not make a correct prediction for the masses of the $\eta$ and $\eta'$ mesons: their values are underestimated by about $300$ MeV. The root of this issue lies in the fact that the matrix model predicts a different quark content for the $\eta$ and $\eta'$ mesons from the Standard Model. To elaborate, the quark model predicts the particles
\begin{equation}
\eta_8=\frac{1}{\sqrt{6}}(u\bar{u}+d\bar{d}-2s\bar{s})
\end{equation}
and
\begin{equation}
\eta_1=\frac{1}{\sqrt{3}}(u\bar{u}+d\bar{d}+s\bar{s}).
\end{equation}
The $\eta$ and $\eta'$ mesons are expressed as the linear combination
\begin{equation}
\left(\begin{array}{c}
			\eta \\
			\eta' \end{array} \right) = \left(\begin{array}{cccc}
\cos\theta_P & -\sin\theta_P \\ 
\sin\theta_P & \cos\theta_P  
\end{array}  \right)
			\left(\begin{array}{c}
			\eta_8 \\
			\eta_1 \end{array} \right)
\end{equation}
where $\theta_P\sim 13.3^{\circ}$ is the mixing angle \cite{Pham:2015ina}.
The mass eigenstates corresponding to $\eta,\eta'$ in our matrix model, however, have a different quark content, which is roughly
\begin{equation}
\eta_{mm}=\frac{1}{\sqrt{2}}(u\bar{u}+d\bar{d}),\quad \eta'_{mm}=s\bar{s}.
\label{etamm}
\end{equation}
This discrepancy occurs because the matrix model seems to be insensitive to the effect of the $U(1)_A$ axial anomaly \cite{Weinberg,Witten,Veneziano} at the current level of approximation. As stated in \cite{Bass, Feldmann}, if the axial anomaly-induced effective gluon mass term is too small, the eigenstates of the mass matrix correspond to hypothetical particles $\eta_q$ and $\eta_s$, which have the same quark content as $\eta_{mm}$ and $\eta'_{mm}$ respectively in (\ref{etamm}), their masses being
\begin{align}
&M_q=m_\pi,\\
&M_s=\sqrt{2 m_K^2-m_\pi^2},
\label{mqs}
\end{align}
where $m_\pi$ and $m_K$ are the masses of the pion and kaon respectively. Therefore, the masses of the $\eta$ and $\eta'$ meson predicted from the matrix model should really be compared with the masses of the hypothetical $\eta_q$ and $\eta_s$ mesons from (\ref{mqs}). This comparison is given in Table \ref{tabeta} below.

\begin{table}[!hbtp]
\centering
\begin{tabular}{|c|c|c|c|c|c|}
\hline
Particle & Spin & Isospin & Matrix Model Mass (MeV) & Observed Mass (MeV) & $\%$ Difference\\ 
\hline
 $\eta_q$ & 0 & 0 & 138.03$\pm$37.88 & 138.04 & 0.0 $\%$ \\ 
\hline 
$\eta_s$& 0 & 0 & 524.14$\pm$70.23 & 687.23 & -23.7 $\%$ \\ 
\hline  
\end{tabular} 
\caption{Comparison of the $\eta$ and $\eta'$ masses in the matrix model with the $\eta_q$ and $\eta_s$ masses.}
\label{tabeta}
\end{table}

 This is an issue that requires further investigation and for this reason, we have excluded  the $\eta$ and $\eta'$ meson in the calculation of $\chi^2$ in Table \ref{T1}.

At this level of approximation, we have considered variational states with up to only $4$ bosonic oscillators and $4$ fermions. Our use of resources, both computational and human, has been modest.

 Our estimates are expected to improve if we increase the size of our variational trial set by including states with up to $6$ bosonic oscillators and $7$ fermions. This larger trial set has several thousand states and involves computations with very large matrices. This requires more efficient numerical codes to be developed for the matrix element computation as well as enumeration of states, and is the focus of our future research.

\begin{figure}[!hbtp]
\centering
\includegraphics[scale=0.5]{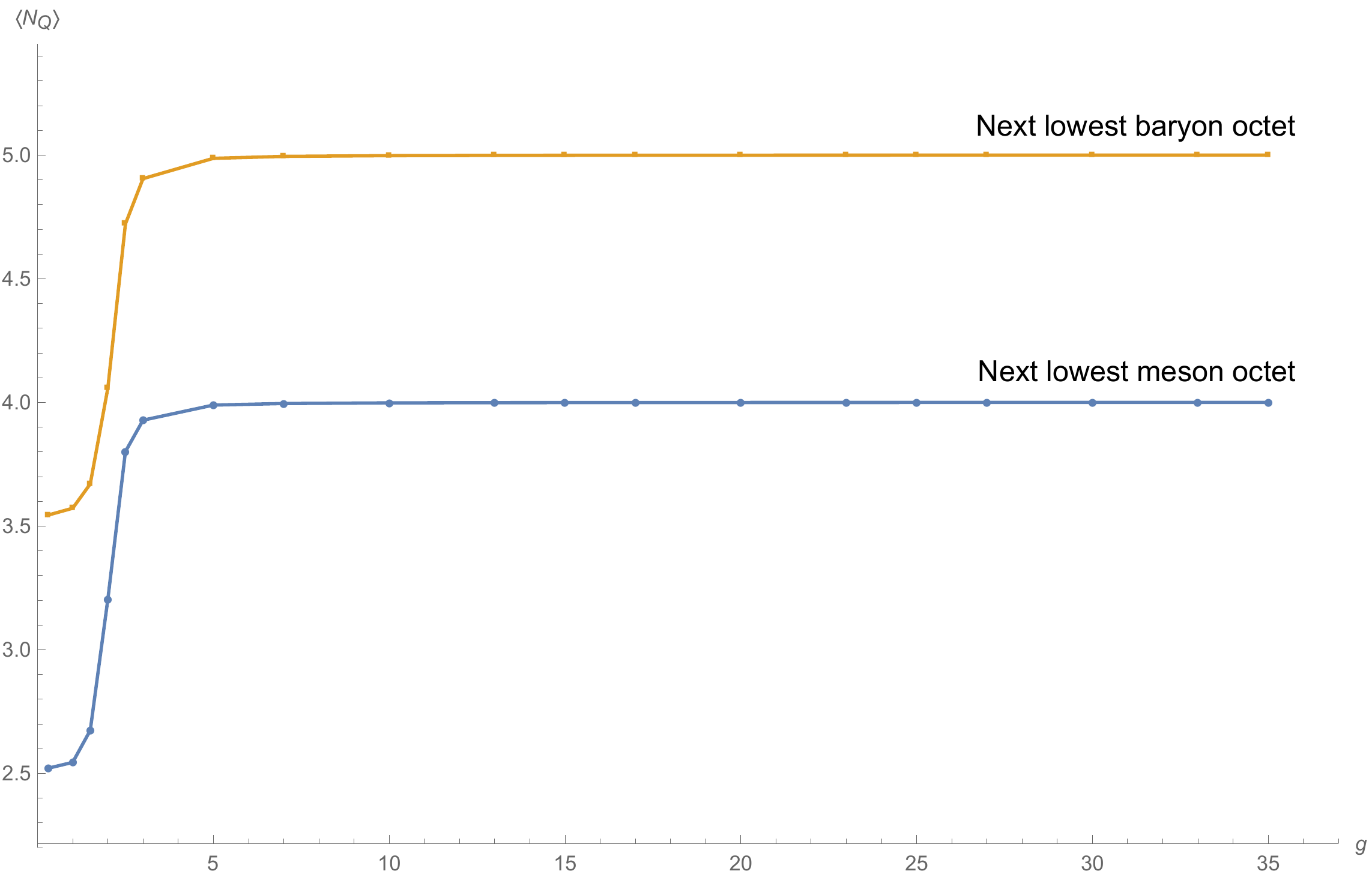}
\caption{A plot of $\langle N_Q\rangle$ vs $g$ for the next lowest meson and baryon octet}
\label{fignq3}
\end{figure}

The next lowest flavour multiplet in the mesonic sector has $\langle N_Q\rangle=3.9999$, and in the baryonic sector has $\langle N_Q\rangle=4.9999$, as seen from Fig. \ref{fignq3}. These are candidates for light exotic hadrons, i.e., hadrons with more than three quarks \cite{Jaffe:2005zz}. These are tetraquark and pentaquark states, with quark content $qq\bar{q}\bar{q}$ and $qqqq\bar{q}$ respectively. Light tetraquarks were first discussed by Jaffe \cite{Jaffe:1976ig} in the context of the bag model \cite{bag1,DeGrand:1975cf}, where he suggested the existence of a tetraquark nonet with mass below 1 GeV, and proposed this as a model for the light scalar mesons. The matrix model prediction for the masses of these particles are about 2.5 GeV. This estimate is expected to improve considerably when we include more states in the trial set.

\appendix
\section{Variational States for the Light Hadrons}\label{varstates}

\subsection{Fermionic states}

We denote the fermion vacuum by $\zf$. Fermionic states can be created by acting on $\zf$ successively by the fermion or the antifermion creation operator $b^\dagger$ and $d^\dagger$ (with the indices suppressed for convenience). A fermionic state can be assigned the quantum numbers $(f,s,c)$, denoting its representation under flavour $SU(3)$, spin and representation under colour $SU(3)$ respectively. A quark state, $b^\dagger\zf$, is denoted by the quantum numbers $\mathbf{(3, \frac{1}{2}, 3)}$ and an antiquark, $d^\dagger\zf$ by $\mathbf{(\bar{3}, \frac{1}{2}, \bar{3})}$.

A "mesonic" state is created by acting on $\zf$ by a pair of quark and antiquark creation operator. Taking a tensor product of the representations gives us all possible ways such a state can transform;
\begin{equation}
\mathbf{(3, \frac{1}{2}, 3)}\otimes \mathbf{(\bar{3}, \frac{1}{2}, \bar{3})}=\mathbf{(8\oplus 1,1\oplus 0, 8\oplus 1)}.
\label{meson}
\end{equation}

This gives the eight possible mesonic states in the ansatz, transforming as \textbf{(1,0,1)}, \textbf{(1,1,1)},
\textbf{(1,0,8)}, \textbf{(1,1,8)}, \textbf{(8,0,1)}, \textbf{(8,1,1)}, \textbf{(8,0,8)} and\textbf{(8,1,8)} respectively.

The interaction term of the Hamiltonian connects a mesonic state to a state with a pair of mesons. Such states can be created by applying on the vacuum two successive meson creation operators. A tensor product decomposition of two mesonic representations gives
\begin{align}
[\mathbf{(3, \frac{1}{2}, 3)}\otimes \mathbf{(\bar{3}, \frac{1}{2}, \bar{3})}]^2&=\mathbf{(8\oplus 1,1\oplus 0, 8\oplus 1)}\otimes \mathbf{(8\oplus 1,1\oplus 0, 8\oplus 1)}\nonumber\\
&=\mathbf{(27\oplus10\oplus\bar{10}\oplus\text{4 }8\oplus \text{2 }1,2\oplus \text{3 }1\oplus \text{2 }0,27\oplus10\oplus\bar{10}\oplus\text{4 }8\oplus \text{2 }1)}.
\label{tens2}
\end{align}

Since the interaction term is a flavour singlet, only those states contribute to a given meson mass whose flavour quantum number is the same as the corresponding meson state. So, we are only interested in the states with $f=1$ or $8$.

The above counting actually gives more states in a particular representation than are actually in the ansatz. For example, looking at $(\ref{tens2}$), there seem to be 16 possible states transforming as $\mathbf{(8,0,1)}$. However, only 4 of them are linearly independent.

A "baryonic" state is created by acting on $\zf$ by three quark creation operators. Taking a tensor product of three quark representations, we find
\begin{equation}
\mathbf{(3, \frac{1}{2}, 3)}\otimes\mathbf{(3, \frac{1}{2}, 3)}\otimes\mathbf{(3, \frac{1}{2}, 3)}=\mathbf{(10\oplus\text{2 }8\oplus 1,\frac{3}{2}\oplus \text{2 }\frac{1}{2}, 10\oplus\text{2 }8\oplus 1)}.
\end{equation}
These states are also organized according to their flavour quantum number. In particular, we organize the states into a baryon decuplet and a baryon octet.

The interaction term $H_{int}$ in (\ref{hint}) connects a baryonic state to a state created by applying on $\zf$ a baryon and a meson creation operator successively. Again, a naive counting by taking the tensor product of representations gives many more states than are linearly independent.

A list of all the linearly independent fermionic states used to create the variational ansatz for light hadrons is given below. Fermionic states are organized according to their quantum numbers $(f,s,c)$. According to the quark content, the states are referred to as mesonic, di-mesonic, baryonic and meson-baryon states. Generators of rotation are denoted by $\tau_i=\frac{1}{2}\sigma_i$ and generators of colour by $T_a=\frac{1}{2}\lambda_a$. Also, generators of the flavour symmetry are denoted by $\TT_F=\frac{1}{2}\lambda_F$.

For notational convenience, spin-2 states are written with two free adjoint indices $i,j$. A spin-2 state can be constructed from a state $|\psi_{ij}\rangle$ with two free adjoint spin indices as
\begin{equation}
|\psi^{(2)}_{ij}\rangle=\frac{1}{2}\left(|\psi_{ij}\rangle+|\psi_{ji}\rangle\right)-\frac{1}{3}\delta_{ij}|\psi_{kk}\rangle
\end{equation}
Similarly, a state transforming in the representation $\mathbf{27},\mathbf{10}$ or $\mathbf{\bar{10}}$ under colour is also written with two free adjoint tensors. The states with definite colour can be constructed as follows;
\begin{equation}
|\psi^{(27)}_{ab}\rangle=\frac{1}{2}\left(|\psi_{ab}\rangle+|\psi_{ba}\rangle\right)-\frac{3}{5}(d_{cde}|\psi_{de}\rangle)d_{abc}-\frac{1}{8}\delta_{ab}|\psi_{cc}\rangle,
\end{equation}
\begin{equation}
|\psi^{(20)}_{ab}\rangle=\frac{1}{2}\left(|\psi_{ab}\rangle-|\psi_{ba}\rangle\right)-\frac{1}{3}(f_{cde}|\psi_{de}\rangle)f_{abc}
\end{equation}
Note that since the $\mathbf{10}$ and $\mathbf{\bar{10}}$ representations are conjugate to each other and orthogonal, they can be combined into a $\mathbf{20}\equiv \mathbf{10\oplus \bar{10}}$.

Meson and di-meson states are organized under flavour $SU(3)$ as an octet or a singlet.
\subsubsection*{Meson States}

\begin{enumerate}
\item \textbf{(1,0,1)}: $b^\dagger_{f \alpha A} (\epsilon c)^\dagger_{f \alpha A}\zf $

\item \textbf{(1,1,1)}: $b^\dagger_{f \alpha A} (\tau_i)_{\alpha\beta} (\epsilon c)^\dagger_{f \beta A}\zf $

\item \textbf{(1,0,8)}: $b^\dagger_{f \alpha A} (T_a)_{AB} (\epsilon c)^\dagger_{f \alpha B}\zf $

\item \textbf{(1,1,8)}: $b^\dagger_{f \alpha A} (\tau_i)_{\alpha\beta} (T_a)_{AB} (\epsilon c)^\dagger_{f \beta B}\zf $

\item \textbf{(8,0,1)}: $(\TT_F)_{fg} b^\dagger_{f \alpha A} (\epsilon c)^\dagger_{g \alpha A}\zf $

\item \textbf{(8,1,1)}: $(\TT_F)_{fg}b^\dagger_{f \alpha A} (\tau_i)_{\alpha\beta} (\epsilon c)^\dagger_{g \beta A}\zf $

\item \textbf{(8,0,8)}: $(\TT_F)_{fg}b^\dagger_{f \alpha A} (T_a)_{AB} (\epsilon c)^\dagger_{g \alpha B}\zf $

\item \textbf{(8,1,8)}: $(\TT_F)_{fg}b^\dagger_{f \alpha A} (\tau_i)_{\alpha\beta} (T_a)_{AB} (\epsilon c)^\dagger_{g \beta B}\zf $

\end{enumerate}

\subsubsection*{Di-Meson States}

\begin{enumerate}
\item \textbf{(1,0,1)}:

\begin{enumerate}
\item $ b^\dagger_{f_1 \alpha_1 A_1} (\epsilon c)^\dagger_{f_1 \alpha_1 A_1}b^\dagger_{f_2 \alpha_2 A_2} (\epsilon c)^\dagger_{f_2 \alpha_2 A_2}|0_F\rangle $

\item $b^\dagger_{f_1 \alpha_1 A_1} (\tau_i)_{\alpha_1\beta_1} (\epsilon c)^\dagger_{f_1 \beta_1 A_1} b^\dagger_{f_2 \alpha_2 A_2} (\tau_i)_{\alpha_2\beta_2} (\epsilon c)^\dagger_{f_2 \beta_2 A_2}\zf$

\item $b^\dagger_{f_1 \alpha_1 A_1} (T_a)_{A_1 B_1} (\epsilon c)^\dagger_{f_1 \alpha_1 B_1} b^\dagger_{f_2 \alpha_2 A_2} (T_a)_{A_2 B_2} (\epsilon c)^\dagger_{f_2 \alpha_2 A_2}\zf$

\item $b^\dagger_{f_1 \alpha_1 A_1} (\tau_i)_{\alpha_1\beta_1} (T_a)_{A_1B_1} (\epsilon c)^\dagger_{f_1 \beta_1 B_1} b^\dagger_{f_2 \alpha_2 A_2} (\tau_i)_{\alpha_2\beta_2} (T_a)_{A_2B_2} (\epsilon c)^\dagger_{f_2 \beta_2 B_2}\zf$
\end{enumerate} 

\item \textbf{(1,1,1)}:
\begin{enumerate}
\item $b^\dagger_{f_1 \alpha_1 A_1} (\tau_i)_{\alpha_1\beta_1} (\epsilon c)^\dagger_{f_1 \beta_1 A_1} b^\dagger_{f_2 \alpha_2 A_2} (\epsilon c)^\dagger_{f_2 \alpha_2 A_2}\zf$

\item $b^\dagger_{f_1 \alpha_1 A_1} (\tau_i)_{\alpha_1\beta_1} (T_a)_{A_1B_1} (\epsilon c)^\dagger_{f_1 \beta_1 B_1} b^\dagger_{f_2 \alpha_2 A_2}(T_a)_{A_2B_2} (\epsilon c)^\dagger_{f_2 \alpha_2 A_2}\zf$
\end{enumerate}

\item \textbf{(1,2,1)}:
\begin{enumerate}
\item $b^\dagger_{f_1 \alpha_1 A_1} (\tau_i)_{\alpha_1\beta_1} (\epsilon c)^\dagger_{f_1 \beta_1 A_1} b^\dagger_{f_2 \alpha_2 A_2} (\tau_j)_{\alpha_2\beta_2}(\epsilon c)^\dagger_{f_2 \beta_2 A_2}\zf$

\item $b^\dagger_{f_1 \alpha_1 A_1} (\tau_i)_{\alpha_1\beta_1} (T_a)_{A_1B_1} (\epsilon c)^\dagger_{f_1 \beta_1 B_1} b^\dagger_{f_2 \alpha_2 A_2}(\tau_j)_{\alpha_2\beta_2}(T_a)_{A_2B_2} (\epsilon c)^\dagger_{f_2 \beta_2 A_2}\zf$
\end{enumerate}

\item \textbf{(1,0,8)}:
\begin{enumerate}
\item $ b^\dagger_{f_1 \alpha_1 A_1} (\epsilon c)^\dagger_{f_1 \alpha_1 A_1}b^\dagger_{f_2 \alpha_2 A_2}(T_a)_{A_2B_2} (\epsilon c)^\dagger_{f_2 \alpha_2 B_2}|0_F\rangle $

\item $d_{abc} b^\dagger_{f_1 \alpha_1 A_1} (T_b)_{A_1B_1} (\epsilon c)^\dagger_{f_1 \alpha_1 B_1}b^\dagger_{f_2 \alpha_2 A_2}(T_c)_{A_2B_2} (\epsilon c)^\dagger_{f_2 \alpha_2 B_2}|0_F\rangle $

\item $ b^\dagger_{f_1 \alpha_1 A_1} (\tau_i)_{\alpha_1\beta_1} (\epsilon c)^\dagger_{f_1 \beta_1 A_1}b^\dagger_{f_2 \alpha_2 A_2}(\tau_i)_{\alpha_2\beta_2}(T_a)_{A_2B_2} (\epsilon c)^\dagger_{f_2 \beta_2 B_2}|0_F\rangle $

\item $d_{abc} b^\dagger_{f_1 \alpha_1 A_1} (\tau_i)_{\alpha_1\beta_1}(T_b)_{A_1B_1} (\epsilon c)^\dagger_{f_1 \beta_1 B_1}b^\dagger_{f_2 \alpha_2 A_2}(\tau_i)_{\alpha_2\beta_2}(T_c)_{A_2B_2} (\epsilon c)^\dagger_{f_2 \beta_2 B_2}|0_F\rangle $

\end{enumerate}

\item \textbf{(1,1,8)}:

\begin{enumerate}

\item $b^\dagger_{f_1 \alpha_1 A_1} (\tau_i)_{\alpha_1\beta_1} (T_a)_{A_1B_1} (\epsilon c)^\dagger_{f_1 \beta_1 B_1} b^\dagger_{f_2 \alpha_2 A_2} (\epsilon c)^\dagger_{f_2 \alpha_2 A_2}\zf $

\item $b^\dagger_{f_1 \alpha_1 A_1} (\tau_i)_{\alpha_1\beta_1}  (\epsilon c)^\dagger_{f_1 \beta_1 A_1} b^\dagger_{f_2 \alpha_2 A_2}(T_a)_{A_2B_2} (\epsilon c)^\dagger_{f_2 \alpha_2 B_2}\zf $

\item $f_{abc}b^\dagger_{f_1 \alpha_1 A_1} (\tau_i)_{\alpha_1\beta_1} (T_b)_{A_1B_1} (\epsilon c)^\dagger_{f_1 \beta_1 B_1} b^\dagger_{f_2 \alpha_2 A_2} (T_c)_{A_2B_2} (\epsilon c)^\dagger_{f_2 \alpha_2 A_2}\zf $

\item $d_{abc}b^\dagger_{f_1 \alpha_1 A_1} (\tau_i)_{\alpha_1\beta_1} (T_b)_{A_1B_1} (\epsilon c)^\dagger_{f_1 \beta_1 B_1} b^\dagger_{f_2 \alpha_2 A_2} (T_c)_{A_2B_2} (\epsilon c)^\dagger_{f_2 \alpha_2 A_2}\zf $

\item $\epsilon_{ijk} b^\dagger_{f_1 \alpha_1 A_1} (\tau_j)_{\alpha_1\beta_1} (T_a)_{A_1B_1} (\epsilon c)^\dagger_{f_1 \beta_1 B_1} b^\dagger_{f_2 \alpha_2 A_2} (\tau_k)_{\alpha_2\beta_2} (\epsilon c)^\dagger_{f_2 \beta_2 A_2}\zf $

\item $\epsilon_{ijk} f_{abc} b^\dagger_{f_1 \alpha_1 A_1} (\tau_j)_{\alpha_1\beta_1} (T_b)_{A_1B_1} (\epsilon c)^\dagger_{f_1 \beta_1 B_1} b^\dagger_{f_2 \alpha_2 A_2} (\tau_k)_{\alpha_2\beta_2} (T_c)_{A_2B_2} (\epsilon c)^\dagger_{f_2 \beta_2 B_2}\zf $
\end{enumerate}

\item \textbf{(1,2,8)}:
\begin{enumerate}
\item $b^\dagger_{f_1 \alpha_1 A_1} (\tau_i)_{\alpha_1\beta_1}(T_a)_{A_1B_1} (\epsilon c)^\dagger_{f_1 \beta_1 B_1} b^\dagger_{f_2 \alpha_2 A_2} (\tau_j)_{\alpha_2\beta_2}(\epsilon c)^\dagger_{f_2 \beta_2 A_2}\zf$

\item $d_{abc}b^\dagger_{f_1 \alpha_1 A_1} (\tau_i)_{\alpha_1\beta_1} (T_b)_{A_1B_1} (\epsilon c)^\dagger_{f_1 \beta_1 B_1} b^\dagger_{f_2 \alpha_2 A_2}(\tau_j)_{\alpha_2\beta_2}(T_c)_{A_2B_2} (\epsilon c)^\dagger_{f_2 \beta_2 A_2}\zf$
\end{enumerate}

\item \textbf{(1,0,27$\oplus$20)}:

\begin{enumerate}
\item $ b^\dagger_{f_1 \alpha_1 A_1} (T_a)_{A_1B_1} (\epsilon c)^\dagger_{f_1 \alpha_1 B_1}b^\dagger_{f_2 \alpha_2 A_2}(T_b)_{A_2B_2} (\epsilon c)^\dagger_{f_2 \alpha_2 B_2}|0_F\rangle $

\item $b^\dagger_{f_1 \alpha_1 A_1} (\tau_i)_{\alpha_1\beta_1} (T_a)_{A_1B_1}(\epsilon c)^\dagger_{f_1 \beta_1 A_1} b^\dagger_{f_2 \alpha_2 A_2} (\tau_i)_{\alpha_2\beta_2} (T_b)_{A_1B_2}(\epsilon c)^\dagger_{f_2 \beta_2 B_2}\zf$
\end{enumerate} 

\item \textbf{(1,1,27$\oplus$20)}:

\begin{enumerate}
\item $ b^\dagger_{f_1 \alpha_1 A_1} (\tau_i)_{\alpha_1\beta_1} (T_a)_{A_1B_1} (\epsilon c)^\dagger_{f_1 \beta_1 B_1}b^\dagger_{f_2 \alpha_2 A_2}(T_b)_{A_2B_2} (\epsilon c)^\dagger_{f_2 \alpha_2 B_2}|0_F\rangle $

\item $\epsilon_{ijk} b^\dagger_{f_1 \alpha_1 A_1} (\tau_j)_{\alpha_1\beta_1} (T_a)_{A_1B_1}(\epsilon c)^\dagger_{f_1 \beta_1 A_1} b^\dagger_{f_2 \alpha_2 A_2} (\tau_k)_{\alpha_2\beta_2} (T_b)_{A_1B_2}(\epsilon c)^\dagger_{f_2 \beta_2 B_2}\zf$
\end{enumerate} 

\item \textbf{(1,2,27$\oplus$20)}:

\begin{enumerate}

\item $b^\dagger_{f_1 \alpha_1 A_1} (\tau_i)_{\alpha_1\beta_1} (T_a)_{A_1B_1}(\epsilon c)^\dagger_{f_1 \beta_1 A_1} b^\dagger_{f_2 \alpha_2 A_2} (\tau_j)_{\alpha_2\beta_2} (T_b)_{A_1B_2}(\epsilon c)^\dagger_{f_2 \beta_2 B_2}\zf$
\end{enumerate} 

\item \textbf{(8,0,1)}:
\begin{enumerate}
\item $ b^\dagger_{f_1 \alpha_1 A_1} (\TT_F)_{f_1g_1} (\epsilon c)^\dagger_{g_1 \alpha_1 A_1}b^\dagger_{f_2 \alpha_2 A_2} (\epsilon c)^\dagger_{f_2 \alpha_2 A_2}\zf $

\item $b^\dagger_{f_1 \alpha_1 A_1}(\TT_F)_{f_1g_1} (\tau_i)_{\alpha_1\beta_1} (\epsilon c)^\dagger_{g_1 \beta_1 A_1} b^\dagger_{f_2 \alpha_2 A_2} (\tau_i)_{\alpha_2\beta_2} (\epsilon c)^\dagger_{f_2 \beta_2 A_2}\zf$

\item $b^\dagger_{f_1 \alpha_1 A_1}(\TT_F)_{f_1g_1} (T_a)_{A_1 B_1} (\epsilon c)^\dagger_{g_1 \alpha_1 B_1} b^\dagger_{f_2 \alpha_2 A_2} (T_a)_{A_2 B_2} (\epsilon c)^\dagger_{f_2 \alpha_2 A_2}\zf$

\item $b^\dagger_{f_1 \alpha_1 A_1} (\TT_F)_{f_1g_1}(\tau_i)_{\alpha_1\beta_1} (T_a)_{A_1B_1} (\epsilon c)^\dagger_{g_1 \beta_1 B_1} b^\dagger_{f_2 \alpha_2 A_2} (\tau_i)_{\alpha_2\beta_2} (T_a)_{A_2B_2} (\epsilon c)^\dagger_{f_2 \beta_2 B_2}\zf$

\end{enumerate}

\item \textbf{(8,1,1)}:
\begin{enumerate}
\item $b^\dagger_{f_1 \alpha_1 A_1}(\TT_F)_{f_1g_1} (\tau_i)_{\alpha_1\beta_1} (\epsilon c)^\dagger_{g_1 \beta_1 A_1} b^\dagger_{f_2 \alpha_2 A_2} (\epsilon c)^\dagger_{f_2 \alpha_2 A_2}\zf$

\item $b^\dagger_{f_1 \alpha_1 A_1}(\TT_F)_{f_1g_1} (\tau_i)_{\alpha_1\beta_1} (T_a)_{A_1B_1} (\epsilon c)^\dagger_{g_1 \beta_1 B_1} b^\dagger_{f_2 \alpha_2 A_2}(T_a)_{A_2B_2} (\epsilon c)^\dagger_{f_2 \alpha_2 A_2}\zf$

\item $b^\dagger_{f_1 \alpha_1 A_1}(\TT_F)_{f_1g_1}  (\epsilon c)^\dagger_{g_1 \alpha_1 A_1} b^\dagger_{f_2 \alpha_2 A_2}(\tau_i)_{\alpha_2\beta_2} (\epsilon c)^\dagger_{f_2 \beta_2 A_2}\zf$

\item $b^\dagger_{f_1 \alpha_1 A_1}(\TT_F)_{f_1g_1}  (T_a)_{A_1B_1} (\epsilon c)^\dagger_{g_1 \alpha_1 B_1} b^\dagger_{f_2 \alpha_2 A_2}(\tau_i)_{\alpha_2\beta_2}(T_a)_{A_2B_2} (\epsilon c)^\dagger_{f_2 \beta_2 A_2}\zf$
\end{enumerate}

\item \textbf{(8,2,1)}:
\begin{enumerate}
\item $b^\dagger_{f_1 \alpha_1 A_1}(\TT_F)_{f_1g_1}  (\tau_i)_{\alpha_1\beta_1} (\epsilon c)^\dagger_{g_1 \beta_1 A_1} b^\dagger_{f_2 \alpha_2 A_2} (\tau_j)_{\alpha_2\beta_2}(\epsilon c)^\dagger_{f_2 \beta_2 A_2}\zf$

\item $b^\dagger_{f_1 \alpha_1 A_1}(\TT_F)_{f_1g_1}  (\tau_i)_{\alpha_1\beta_1} (T_a)_{A_1B_1} (\epsilon c)^\dagger_{g_1 \beta_1 B_1} b^\dagger_{f_2 \alpha_2 A_2}(\tau_j)_{\alpha_2\beta_2}(T_a)_{A_2B_2} (\epsilon c)^\dagger_{f_2 \beta_2 A_2}\zf$
\end{enumerate}

\item \textbf{(8,0,8)}:
\begin{enumerate}
\item $ b^\dagger_{f_1 \alpha_1 A_1}(\TT_F)_{f_1g_1} (T_a)_{A_1B_1} (\epsilon c)^\dagger_{g_1 \alpha_1 B_1}b^\dagger_{f_2 \alpha_2 A_2} (\epsilon c)^\dagger_{f_2 \alpha_2 A_2}|0_F\rangle $
\item $ b^\dagger_{f_1 \alpha_1 A_1}(\TT_F)_{f_1g_1}  (\tau_i)_{\alpha_1\beta_1} (T_a)_{A_1B_1}(\epsilon c)^\dagger_{g_1 \beta_1 B_1}b^\dagger_{f_2 \alpha_2 A_2}(\tau_i)_{\alpha_2\beta_2} (\epsilon c)^\dagger_{f_2 \beta_2 A_2}|0_F\rangle $

\item $ b^\dagger_{f_1 \alpha_1 A_1}(\TT_F)_{f_1g_1}  (\epsilon c)^\dagger_{g_1 \alpha_1 A_1}b^\dagger_{f_2 \alpha_2 A_2}(T_a)_{A_2B_2} (\epsilon c)^\dagger_{f_2 \alpha_2 B_2}|0_F\rangle $
\item $ b^\dagger_{f_1 \alpha_1 A_1}(\TT_F)_{f_1g_1}  (\tau_i)_{\alpha_1\beta_1} (\epsilon c)^\dagger_{g_1 \beta_1 A_1}b^\dagger_{f_2 \alpha_2 A_2}(\tau_i)_{\alpha_2\beta_2}(T_a)_{A_2B_2} (\epsilon c)^\dagger_{f_2 \beta_2 B_2}|0_F\rangle $

\item $f_{abc} b^\dagger_{f_1 \alpha_1 A_1}(\TT_F)_{f_1g_1}  (T_b)_{A_1B_1} (\epsilon c)^\dagger_{g_1 \alpha_1 B_1}b^\dagger_{f_2 \alpha_2 A_2}(T_c)_{A_2B_2} (\epsilon c)^\dagger_{f_2 \alpha_2 B_2}|0_F\rangle $
\item $d_{abc} b^\dagger_{f_1 \alpha_1 A_1}(\TT_F)_{f_1g_1}  (T_b)_{A_1B_1} (\epsilon c)^\dagger_{g_1 \alpha_1 B_1}b^\dagger_{f_2 \alpha_2 A_2}(T_c)_{A_2B_2} (\epsilon c)^\dagger_{f_2 \alpha_2 B_2}|0_F\rangle $

\item $f_{abc} b^\dagger_{f_1 \alpha_1 A_1}(\TT_F)_{f_1g_1}  (\tau_i)_{\alpha_1\beta_1}(T_b)_{A_1B_1} (\epsilon c)^\dagger_{g_1 \beta_1 B_1}b^\dagger_{f_2 \alpha_2 A_2}(\tau_i)_{\alpha_2\beta_2}(T_c)_{A_2B_2} (\epsilon c)^\dagger_{f_2 \beta_2 B_2}|0_F\rangle $
\item $d_{abc} b^\dagger_{f_1 \alpha_1 A_1}(\TT_F)_{f_1g_1}  (\tau_i)_{\alpha_1\beta_1}(T_b)_{A_1B_1} (\epsilon c)^\dagger_{g_1 \beta_1 B_1}b^\dagger_{f_2 \alpha_2 A_2}(\tau_i)_{\alpha_2\beta_2}(T_c)_{A_2B_2} (\epsilon c)^\dagger_{f_2 \beta_2 B_2}|0_F\rangle $

\end{enumerate}

\item \textbf{(8,1,8)}:

\begin{enumerate}
\item $b^\dagger_{f_1 \alpha_1 A_1} (\TT_F)_{f_1g_1} (\tau_i)_{\alpha_1\beta_1} (T_a)_{A_1B_1} (\epsilon c)^\dagger_{g_1 \beta_1 B_1} b^\dagger_{f_2 \alpha_2 A_2} (\epsilon c)^\dagger_{f_2 \alpha_2 A_2}\zf $

\item $b^\dagger_{f_1 \alpha_1 A_1}(\TT_F)_{f_1g_1} (\tau_i)_{\alpha_1\beta_1}  (\epsilon c)^\dagger_{g_1 \beta_1 A_1} b^\dagger_{f_2 \alpha_2 A_2}(T_a)_{A_2B_2} (\epsilon c)^\dagger_{f_2 \alpha_2 B_2}\zf $

\item $b^\dagger_{f_1 \alpha_1 A_1} (\TT_F)_{f_1g_1} (T_a)_{A_1B_1} (\epsilon c)^\dagger_{g_1 \alpha_1 B_1} b^\dagger_{f_2 \alpha_2 A_2}(\tau_i)_{\alpha_2\beta_2} (\epsilon c)^\dagger_{f_2 \beta_2 A_2}\zf $

\item $b^\dagger_{f_1 \alpha_1 A_1}(\TT_F)_{f_1g_1}   (\epsilon c)^\dagger_{g_1 \alpha_1 A_1} b^\dagger_{f_2 \alpha_2 A_2}(\tau_i)_{\alpha_2\beta_2}(T_a)_{A_2B_2} (\epsilon c)^\dagger_{f_2 \beta_2 B_2}\zf $

\item $f_{abc}b^\dagger_{f_1 \alpha_1 A_1}(\TT_F)_{f_1g_1} (\tau_i)_{\alpha_1\beta_1} (T_b)_{A_1B_1} (\epsilon c)^\dagger_{g_1 \beta_1 B_1} b^\dagger_{f_2 \alpha_2 A_2} (T_c)_{A_2B_2} (\epsilon c)^\dagger_{f_2 \alpha_2 A_2}\zf $

\item $d_{abc}b^\dagger_{f_1 \alpha_1 A_1}(\TT_F)_{f_1g_1} (\tau_i)_{\alpha_1\beta_1} (T_b)_{A_1B_1} (\epsilon c)^\dagger_{g_1 \beta_1 B_1} b^\dagger_{f_2 \alpha_2 A_2} (T_c)_{A_2B_2} (\epsilon c)^\dagger_{f_2 \alpha_2 A_2}\zf $

\item $f_{abc}b^\dagger_{f_1 \alpha_1 A_1}(\TT_F)_{f_1g_1} (T_b)_{A_1B_1} (\epsilon c)^\dagger_{g_1 \alpha_1 B_1} b^\dagger_{f_2 \alpha_2 A_2}(\tau_i)_{\alpha_2\beta_2} (T_c)_{A_2B_2} (\epsilon c)^\dagger_{f_2 \beta_2 A_2}\zf $

\item $d_{abc}b^\dagger_{f_1 \alpha_1 A_1}(\TT_F)_{f_1g_1} (T_b)_{A_1B_1} (\epsilon c)^\dagger_{g_1 \alpha_1 B_1} b^\dagger_{f_2 \alpha_2 A_2}(\tau_i)_{\alpha_2\beta_2} (T_c)_{A_2B_2} (\epsilon c)^\dagger_{f_2 \beta_2 A_2}\zf $

\item $\epsilon_{ijk} b^\dagger_{f_1 \alpha_1 A_1}(\TT_F)_{f_1g_1} (\tau_j)_{\alpha_1\beta_1} (T_a)_{A_1B_1} (\epsilon c)^\dagger_{g_1 \beta_1 B_1} b^\dagger_{f_2 \alpha_2 A_2} (\tau_k)_{\alpha_2\beta_2} (\epsilon c)^\dagger_{f_2 \beta_2 A_2}\zf $

\item $\epsilon_{ijk} b^\dagger_{f_1 \alpha_1 A_1}(\TT_F)_{f_1g_1} (\tau_j)_{\alpha_1\beta_1} (\epsilon c)^\dagger_{g_1 \beta_1 A_1} b^\dagger_{f_2 \alpha_2 A_2} (\tau_k)_{\alpha_2\beta_2} (T_a)_{A_2B_2}(\epsilon c)^\dagger_{f_2 \beta_2 B_2}\zf $

\item $\epsilon_{ijk} f_{abc} b^\dagger_{f_1 \alpha_1 A_1} (\TT_F)_{f_1g_1}(\tau_j)_{\alpha_1\beta_1} (T_b)_{A_1B_1} (\epsilon c)^\dagger_{g_1 \beta_1 B_1} b^\dagger_{f_2 \alpha_2 A_2} (\tau_k)_{\alpha_2\beta_2} (T_c)_{A_2B_2} (\epsilon c)^\dagger_{f_2 \beta_2 B_2}\zf $
\item $\epsilon_{ijk} d_{abc} b^\dagger_{f_1 \alpha_1 A_1} (\TT_F)_{f_1g_1}(\tau_j)_{\alpha_1\beta_1} (T_b)_{A_1B_1} (\epsilon c)^\dagger_{g_1 \beta_1 B_1} b^\dagger_{f_2 \alpha_2 A_2} (\tau_k)_{\alpha_2\beta_2} (T_c)_{A_2B_2} (\epsilon c)^\dagger_{f_2 \beta_2 B_2}\zf $
\end{enumerate}

\item \textbf{(8,2,8)}:
\begin{enumerate}
\item $b^\dagger_{f_1 \alpha_1 A_1}(\TT_F)_{f_1g_1} (\tau_i)_{\alpha_1\beta_1}(T_a)_{A_1B_1} (\epsilon c)^\dagger_{g_1 \beta_1 B_1} b^\dagger_{f_2 \alpha_2 A_2} (\tau_j)_{\alpha_2\beta_2}(\epsilon c)^\dagger_{f_2 \beta_2 A_2}\zf$

\item $b^\dagger_{f_1 \alpha_1 A_1}(\TT_F)_{f_1g_1} (\tau_i)_{\alpha_1\beta_1} (\epsilon c)^\dagger_{g_1 \beta_1 A_1} b^\dagger_{f_2 \alpha_2 A_2} (\tau_j)_{\alpha_2\beta_2}(T_a)_{A_2B_2}(\epsilon c)^\dagger_{f_2 \beta_2 B_2}\zf$

\item $f_{abc}b^\dagger_{f_1 \alpha_1 A_1}(\TT_F)_{f_1g_1} (\tau_i)_{\alpha_1\beta_1} (T_b)_{A_1B_1} (\epsilon c)^\dagger_{g_1 \beta_1 B_1} b^\dagger_{f_2 \alpha_2 A_2}(\tau_j)_{\alpha_2\beta_2}(T_c)_{A_2B_2} (\epsilon c)^\dagger_{f_2 \beta_2 A_2}\zf$

\item $d_{abc}b^\dagger_{f_1 \alpha_1 A_1}(\TT_F)_{f_1g_1} (\tau_i)_{\alpha_1\beta_1} (T_b)_{A_1B_1} (\epsilon c)^\dagger_{g_1 \beta_1 B_1} b^\dagger_{f_2 \alpha_2 A_2}(\tau_j)_{\alpha_2\beta_2}(T_c)_{A_2B_2} (\epsilon c)^\dagger_{f_2 \beta_2 A_2}\zf$
\end{enumerate}

\item \textbf{(8,0,27$\oplus$20)}:

\begin{enumerate}
\item $ b^\dagger_{f_1 \alpha_1 A_1} (\TT_F)_{f_1g_1} (T_a)_{A_1B_1} (\epsilon c)^\dagger_{g_1 \alpha_1 B_1}b^\dagger_{f_2 \alpha_2 A_2}(T_b)_{A_2B_2} (\epsilon c)^\dagger_{f_2 \alpha_2 B_2}|0_F\rangle $

\item $b^\dagger_{f_1 \alpha_1 A_1}(\TT_F)_{f_1g_1}  (\tau_i)_{\alpha_1\beta_1} (T_a)_{A_1B_1}(\epsilon c)^\dagger_{g_1 \beta_1 A_1} b^\dagger_{f_2 \alpha_2 A_2} (\tau_i)_{\alpha_2\beta_2} (T_b)_{A_1B_2}(\epsilon c)^\dagger_{f_2 \beta_2 B_2}\zf$
\end{enumerate} 

\item \textbf{(8,1,27$\oplus$20)}:

\begin{enumerate}
\item $ b^\dagger_{f_1 \alpha_1 A_1}(\TT_F)_{f_1g_1}  (\tau_i)_{\alpha_1\beta_1} (T_a)_{A_1B_1} (\epsilon c)^\dagger_{g_1 \beta_1 B_1}b^\dagger_{f_2 \alpha_2 A_2}(T_b)_{A_2B_2} (\epsilon c)^\dagger_{f_2 \alpha_2 B_2}|0_F\rangle $

\item $\epsilon_{ijk} b^\dagger_{f_1 \alpha_1 A_1}(\TT_F)_{f_1g_1}  (\tau_j)_{\alpha_1\beta_1} (T_a)_{A_1B_1}(\epsilon c)^\dagger_{g_1 \beta_1 A_1} b^\dagger_{f_2 \alpha_2 A_2} (\tau_k)_{\alpha_2\beta_2} (T_b)_{A_1B_2}(\epsilon c)^\dagger_{f_2 \beta_2 B_2}\zf$
\end{enumerate} 

\item \textbf{(8,2,27$\oplus$20)}:

\begin{enumerate}

\item $b^\dagger_{f_1 \alpha_1 A_1}(\TT_F)_{f_1g_1}  (\tau_i)_{\alpha_1\beta_1} (T_a)_{A_1B_1}(\epsilon c)^\dagger_{g_1 \beta_1 A_1} b^\dagger_{f_2 \alpha_2 A_2} (\tau_j)_{\alpha_2\beta_2} (T_b)_{A_1B_2}(\epsilon c)^\dagger_{f_2 \beta_2 B_2}\zf$
\end{enumerate} 

\end{enumerate}

Baryon and baryon-meson states are organized into an octet and a decuplet under flavour $SU(3)$. We define $(T_F)_{fgh}\equiv (\TT_F)_{ff_1}\epsilon_{f_1gh}$, $f,g,h=1,...,3$ and $F=1,...8$. We also define the tensor $\Delta^\rho_{f,g,h}$, $f,g,h=1,...,3$ and $\rho=1,...,10$ such that $\Delta$ is totally symmetric in $f,g,h$ and generates the flavour decuplet. States that carry both adjoint spin indices $i,j,...$ and a fundamental spin index $\gamma$ can be separated into different spin components using suitable projectors, however it is convenient to express them like this before combining them with gluon wavefunctions. It is also convenient to define $(\mathcal{E}_A)_{BC}\equiv \epsilon_{ABC}$ for fundamental indices $A,B,C=1$ to $3$.

\subsubsection*{Baryon States}

\begin{enumerate}
\item $\mathbf{(8,\frac{1}{2},1)}:(T_F)_{fgh}(\EE_A)_{BC}\epsilon_{\alpha\beta} b^\dagger_{f \alpha A} b^\dagger_{g \beta B}b^\dagger_{h \gamma C}\zf $

\item  $\mathbf{(8,1\otimes \frac{1}{2},1)}:(T_F)_{fgh}(\EE_A)_{BC}(\tau_i\epsilon)_{\alpha\beta}b^\dagger_{f \alpha A} b^\dagger_{g \beta B}b^\dagger_{h \gamma C}\zf$

\item  $\mathbf{(8,\frac{1}{2},8)}:(T_F)_{fgh}(\EE_A T_a)_{BC}\epsilon_{\alpha\beta}b^\dagger_{f \alpha A} b^\dagger_{g \beta B}b^\dagger_{h \gamma C}\zf $

\item  $\mathbf{(8,1\otimes \frac{1}{2},8)}:(T_F)_{fgh}(\EE_A T_a)_{BC}(\tau_i\epsilon)_{\alpha\beta}b^\dagger_{f \alpha A} b^\dagger_{g \beta B}b^\dagger_{h \gamma C}\zf $


\item  $\mathbf{(10,1\otimes \frac{1}{2},1)}:\Delta^\rho_{fgh}(\EE_A)_{BC}(\tau_i\epsilon)_{\alpha\beta}b^\dagger_{f \alpha A} b^\dagger_{g \beta B}b^\dagger_{h \gamma C}\zf$

\item  $\mathbf{(10,\frac{1}{2},8)}:\Delta^\rho_{fgh}(\EE_A T_a)_{BC}\epsilon_{\alpha\beta}b^\dagger_{f \alpha A} b^\dagger_{g \beta B}b^\dagger_{h \gamma C}\zf $

\item  $\mathbf{(10,1\otimes \frac{1}{2},8)}:\Delta^\rho_{fgh}(\EE_A T_a)_{BC}(\tau_i\epsilon)_{\alpha\beta}b^\dagger_{f \alpha A} b^\dagger_{g \beta B}b^\dagger_{h \gamma C}\zf $

\end{enumerate}
\textbf{Note:}$\mathbf{(10,\frac{1}{2},1)}$ does not exist due to symmetry.

\subsubsection*{Baryon-Meson States}

\begin{enumerate}
\item $\mathbf{(8,\frac{1}{2},1)}:$

\begin{enumerate}
\item $ (T_F)_{fgh}(\EE_A)_{BC}\epsilon_{\alpha\beta} b^\dagger_{f \alpha A} b^\dagger_{g \beta B}b^\dagger_{h \gamma C}b^\dagger_{f_2 \alpha_2 A_2} (\epsilon c)^\dagger_{f_2 \alpha_2 A_2}|0_F\rangle $

\item $(T_F)_{fgh}(\EE_A)_{BC}(\tau_i\epsilon)_{\alpha\beta}b^\dagger_{f \alpha A} b^\dagger_{g \beta B}b^\dagger_{h \gamma C} b^\dagger_{f_2 \alpha_2 A_2} (\tau_i)_{\alpha_2\beta_2} (\epsilon c)^\dagger_{f_2 \beta_2 A_2}\zf$

\item $(T_F)_{fgh}(\EE_A T_a)_{BC}\epsilon_{\alpha\beta}b^\dagger_{f \alpha A} b^\dagger_{g \beta B}b^\dagger_{h \gamma C} b^\dagger_{f_2 \alpha_2 A_2} (T_a)_{A_2 B_2} (\epsilon c)^\dagger_{f_2 \alpha_2 A_2}\zf$

\item $(T_F)_{fgh}(\EE_A T_a)_{BC}(\tau_i\epsilon)_{\alpha\beta}b^\dagger_{f \alpha A} b^\dagger_{g \beta B}b^\dagger_{h \gamma C} b^\dagger_{f_2 \alpha_2 A_2} (\tau_i)_{\alpha_2\beta_2} (T_a)_{A_2B_2} (\epsilon c)^\dagger_{f_2 \beta_2 B_2}\zf$
\end{enumerate} 

\item $\mathbf{(8,1\otimes\frac{1}{2},1)}:$
\begin{enumerate}
\item $ (T_F)_{fgh}(\EE_A)_{BC}(\tau_i\epsilon)_{\alpha\beta} b^\dagger_{f \alpha A} b^\dagger_{g \beta B}b^\dagger_{h \gamma C}b^\dagger_{f_2 \alpha_2 A_2} (\epsilon c)^\dagger_{f_2 \alpha_2 A_2}|0_F\rangle $

\item $(T_F)_{fgh}(\EE_A T_a)_{BC}\epsilon_{\alpha\beta}b^\dagger_{f \alpha A} b^\dagger_{g \beta B}b^\dagger_{h \gamma C} b^\dagger_{f_2 \alpha_2 A_2} (\tau_i)_{\alpha_2\beta_2} (T_a)_{A_2B_2} (\epsilon c)^\dagger_{f_2 \beta_2 B_2}\zf$

\item $(T_F)_{fgh}(\EE_A T_a)_{BC}(\tau_i\epsilon)_{\alpha\beta}b^\dagger_{f \alpha A} b^\dagger_{g \beta B}b^\dagger_{h \gamma C} b^\dagger_{f_2 \alpha_2 A_2} (T_a)_{A_2B_2} (\epsilon c)^\dagger_{f_2 \alpha_2 B_2}\zf$

\item $\epsilon_{ijk} (T_F)_{fgh}(\EE_A)_{BC}(\tau_j\epsilon)_{\alpha\beta} b^\dagger_{f \alpha A} b^\dagger_{g \beta B}b^\dagger_{h \gamma C}b^\dagger_{f_2 \alpha_2 A_2}(\tau_k)_{\alpha_2\beta_2} (\epsilon c)^\dagger_{f_2 \beta_2 A_2}|0_F\rangle $

\item $\epsilon_{ijk} (T_F)_{fgh}(\EE_A T_a)_{BC}(\tau_j\epsilon)_{\alpha\beta} b^\dagger_{f \alpha A} b^\dagger_{g \beta B}b^\dagger_{h \gamma C}b^\dagger_{f_2 \alpha_2 A_2}(\tau_k)_{\alpha_2\beta_2} (T_a)_{A_2 B_2} (\epsilon c)^\dagger_{f_2 \beta_2 B_2}|0_F\rangle $

\item $(T_F)_{fgh}(\EE_A)_{BC}\epsilon_{\alpha\beta}b^\dagger_{f \alpha A} b^\dagger_{g \beta B}b^\dagger_{h \gamma C} b^\dagger_{f_2 \alpha_2 A_2} (\tau_i)_{\alpha_2\beta_2} (\epsilon c)^\dagger_{f_2 \beta_2 B_2}\zf$

\end{enumerate} 

\item $\mathbf{(8,2\otimes\frac{1}{2},1)}:$
\begin{enumerate}
\item $(T_F)_{fgh}(\EE_A)_{BC}(\tau_i\epsilon)_{\alpha\beta} b^\dagger_{f \alpha A} b^\dagger_{g \beta B}b^\dagger_{h \gamma C} b^\dagger_{f_2 \alpha_2 A_2} (\tau_j)_{\alpha_2\beta_2}(\epsilon c)^\dagger_{f_2 \beta_2 A_2}\zf$

\item $(T_F)_{fgh}(\EE_A T_a)_{BC}(\tau_i\epsilon)_{\alpha\beta} b^\dagger_{f \alpha A} b^\dagger_{g \beta B}b^\dagger_{h \gamma C} b^\dagger_{f_2 \alpha_2 A_2}(\tau_j)_{\alpha_2\beta_2}(T_a)_{A_2B_2} (\epsilon c)^\dagger_{f_2 \beta_2 A_2}\zf$
\end{enumerate}

\item $\mathbf{(8,\frac{1}{2},8)}:$
\begin{enumerate}
\item $(T_F)_{fgh}(\EE_A)_{BC}\epsilon_{\alpha\beta} b^\dagger_{f \alpha A} b^\dagger_{g \beta B}b^\dagger_{h \gamma C}b^\dagger_{f_2 \alpha_2 A_2}(T_a)_{A_2B_2} (\epsilon c)^\dagger_{f_2 \alpha_2 B_2}|0_F\rangle $
\item $ (T_F)_{fgh}(\EE_A T_a)_{BC}\epsilon_{\alpha\beta} b^\dagger_{f \alpha A} b^\dagger_{g \beta B}b^\dagger_{h \gamma C}b^\dagger_{f_2 \alpha_2 A_2} (\epsilon c)^\dagger_{f_2 \alpha_2 A_2}|0_F\rangle $
\item $ (T_F)_{fgh}(\EE_A)_{BC}(\tau_i\epsilon)_{\alpha\beta} b^\dagger_{f \alpha A} b^\dagger_{g \beta B}b^\dagger_{h \gamma C}b^\dagger_{f_2 \alpha_2 A_2}(\tau_i)_{\alpha_2\beta_2}(T_a)_{A_2B_2} (\epsilon c)^\dagger_{f_2 \beta_2 B_2}|0_F\rangle $
\item $(T_F)_{fgh}(\EE_A T_a)_{BC}(\tau_i\epsilon)_{\alpha\beta} b^\dagger_{f \alpha A} b^\dagger_{g \beta B}b^\dagger_{h \gamma C}b^\dagger_{f_2 \alpha_2 A_2}(\tau_i)_{\alpha_2\beta_2} (\epsilon c)^\dagger_{f_2 \beta_2 A_2}|0_F\rangle $

\item $f_{abc} (T_F)_{fgh}(\EE_A T_b)_{BC}\epsilon_{\alpha\beta} b^\dagger_{f \alpha A} b^\dagger_{g \beta B}b^\dagger_{h \gamma C}b^\dagger_{f_2 \alpha_2 A_2}(T_c)_{A_2B_2} (\epsilon c)^\dagger_{f_2 \alpha_2 B_2}|0_F\rangle $
\item $d_{abc} (T_F)_{fgh}(\EE_A T_b)_{BC}\epsilon_{\alpha\beta} b^\dagger_{f \alpha A} b^\dagger_{g \beta B}b^\dagger_{h \gamma C}b^\dagger_{f_2 \alpha_2 A_2}(T_c)_{A_2B_2} (\epsilon c)^\dagger_{f_2 \alpha_2 B_2}|0_F\rangle $

\item $f_{abc} (T_F)_{fgh}(\EE_A T_b)_{BC}(\tau_i\epsilon)_{\alpha\beta} b^\dagger_{f \alpha A} b^\dagger_{g \beta B}b^\dagger_{h \gamma C}b^\dagger_{f_2 \alpha_2 A_2}(\tau_i)_{\alpha_2\beta_2}(T_c)_{A_2B_2} (\epsilon c)^\dagger_{f_2 \beta_2 B_2}|0_F\rangle $
\item $d_{abc} (T_F)_{fgh}(\EE_A T_b)_{BC}(\tau_i\epsilon)_{\alpha\beta} b^\dagger_{f \alpha A} b^\dagger_{g \beta B}b^\dagger_{h \gamma C}b^\dagger_{f_2 \alpha_2 A_2}(\tau_i)_{\alpha_2\beta_2}(T_c)_{A_2B_2} (\epsilon c)^\dagger_{f_2 \beta_2 B_2}|0_F\rangle $

\end{enumerate}

\item $\mathbf{(8,1\otimes\frac{1}{2},8)}:$

\begin{enumerate}
\item $(T_F)_{fgh}(\EE_A T_a)_{BC}(\tau_i\epsilon)_{\alpha\beta} b^\dagger_{f \alpha A} b^\dagger_{g \beta B}b^\dagger_{h \gamma C} b^\dagger_{f_2 \alpha_2 A_2} (\epsilon c)^\dagger_{f_2 \alpha_2 A_2}\zf $

\item $(T_F)_{fgh}(\EE_A)_{BC}(\tau_i\epsilon)_{\alpha\beta} b^\dagger_{f \alpha A} b^\dagger_{g \beta B}b^\dagger_{h \gamma C} b^\dagger_{f_2 \alpha_2 A_2}(T_a)_{A_2B_2} (\epsilon c)^\dagger_{f_2 \alpha_2 B_2}\zf $

\item $(T_F)_{fgh}(\EE_A T_a)_{BC}\epsilon_{\alpha\beta} b^\dagger_{f \alpha A} b^\dagger_{g \beta B}b^\dagger_{h \gamma C} b^\dagger_{f_2 \alpha_2 A_2}(\tau_i)_{\alpha_2\beta_2} (\epsilon c)^\dagger_{f_2 \beta_2 A_2}\zf $

\item $(T_F)_{fgh}(\EE_A)_{BC}\epsilon_{\alpha\beta} b^\dagger_{f \alpha A} b^\dagger_{g \beta B}b^\dagger_{h \gamma C} b^\dagger_{f_2 \alpha_2 A_2}(\tau_i)_{\alpha_2\beta_2}(T_a)_{A_2B_2} (\epsilon c)^\dagger_{f_2 \beta_2 B_2}\zf $

\item $f_{abc}(T_F)_{fgh}(\EE_A T_b)_{BC}(\tau_i\epsilon)_{\alpha\beta} b^\dagger_{f \alpha A} b^\dagger_{g \beta B}b^\dagger_{h \gamma C} b^\dagger_{f_2 \alpha_2 A_2} (T_c)_{A_2B_2} (\epsilon c)^\dagger_{f_2 \alpha_2 A_2}\zf $

\item $d_{abc}(T_F)_{fgh}(\EE_A T_b)_{BC}(\tau_i\epsilon)_{\alpha\beta} b^\dagger_{f \alpha A} b^\dagger_{g \beta B}b^\dagger_{h \gamma C} b^\dagger_{f_2 \alpha_2 A_2} (T_c)_{A_2B_2} (\epsilon c)^\dagger_{f_2 \alpha_2 A_2}\zf $

\item $f_{abc}(T_F)_{fgh}(\EE_A T_b)_{BC}\epsilon_{\alpha\beta} b^\dagger_{f \alpha A} b^\dagger_{g \beta B}b^\dagger_{h \gamma C} b^\dagger_{f_2 \alpha_2 A_2}(\tau_i)_{\alpha_2\beta_2} (T_c)_{A_2B_2} (\epsilon c)^\dagger_{f_2 \beta_2 A_2}\zf $

\item $d_{abc}(T_F)_{fgh}(\EE_A T_b)_{BC}\epsilon_{\alpha\beta} b^\dagger_{f \alpha A} b^\dagger_{g \beta B}b^\dagger_{h \gamma C} b^\dagger_{f_2 \alpha_2 A_2}(\tau_i)_{\alpha_2\beta_2} (T_c)_{A_2B_2} (\epsilon c)^\dagger_{f_2 \beta_2 A_2}\zf $

\item $\epsilon_{ijk} (T_F)_{fgh}(\EE_A)_{BC}(\tau_j\epsilon)_{\alpha\beta} b^\dagger_{f \alpha A} b^\dagger_{g \beta B}b^\dagger_{h \gamma C} b^\dagger_{f_2 \alpha_2 A_2} (\tau_k)_{\alpha_2\beta_2} (\epsilon c)^\dagger_{f_2 \beta_2 A_2}\zf $

\item $\epsilon_{ijk} (T_F)_{fgh}(\EE_A)_{BC}(\tau_j\epsilon)_{\alpha\beta} b^\dagger_{f \alpha A} b^\dagger_{g \beta B}b^\dagger_{h \gamma C} b^\dagger_{f_2 \alpha_2 A_2} (\tau_k)_{\alpha_2\beta_2} (T_a)_{A_2B_2}(\epsilon c)^\dagger_{f_2 \beta_2 B_2}\zf $

\item $\epsilon_{ijk} f_{abc} (T_F)_{fgh}(\EE_A T_b)_{BC}(\tau_j\epsilon)_{\alpha\beta} b^\dagger_{f \alpha A} b^\dagger_{g \beta B}b^\dagger_{h \gamma C} b^\dagger_{f_2 \alpha_2 A_2} (\tau_k)_{\alpha_2\beta_2} (T_c)_{A_2B_2} (\epsilon c)^\dagger_{f_2 \beta_2 B_2}\zf $

\item $\epsilon_{ijk} d_{abc} (T_F)_{fgh}(\EE_A T_b)_{BC}(\tau_j\epsilon)_{\alpha\beta} b^\dagger_{f \alpha A} b^\dagger_{g \beta B}b^\dagger_{h \gamma C} b^\dagger_{f_2 \alpha_2 A_2} (\tau_k)_{\alpha_2\beta_2} (T_c)_{A_2B_2} (\epsilon c)^\dagger_{f_2 \beta_2 B_2}\zf $
\end{enumerate}

\item $\mathbf{(8,2\otimes\frac{1}{2},8)}:$
\begin{enumerate}
\item $(T_F)_{fgh}(\EE_A T_a)_{BC}(\tau_i\epsilon)_{\alpha\beta} b^\dagger_{f \alpha A} b^\dagger_{g \beta B}b^\dagger_{h \gamma C} b^\dagger_{f_2 \alpha_2 A_2} (\tau_j)_{\alpha_2\beta_2}(\epsilon c)^\dagger_{f_2 \beta_2 A_2}\zf$

\item $(T_F)_{fgh}(\EE_A)_{BC}(\tau_i\epsilon)_{\alpha\beta} b^\dagger_{f \alpha A} b^\dagger_{g \beta B}b^\dagger_{h \gamma C} b^\dagger_{f_2 \alpha_2 A_2} (\tau_j)_{\alpha_2\beta_2}(T_a)_{A_2B_2}(\epsilon c)^\dagger_{f_2 \beta_2 B_2}\zf$

\item $f_{abc}(T_F)_{fgh}(\EE_A T_b)_{BC}(\tau_i\epsilon)_{\alpha\beta} b^\dagger_{f \alpha A} b^\dagger_{g \beta B}b^\dagger_{h \gamma C} b^\dagger_{f_2 \alpha_2 A_2}(\tau_j)_{\alpha_2\beta_2}(T_c)_{A_2B_2} (\epsilon c)^\dagger_{f_2 \beta_2 A_2}\zf$

\item $d_{abc}(T_F)_{fgh}(\EE_A T_b)_{BC}(\tau_i\epsilon)_{\alpha\beta} b^\dagger_{f \alpha A} b^\dagger_{g \beta B}b^\dagger_{h \gamma C} b^\dagger_{f_2 \alpha_2 A_2}(\tau_j)_{\alpha_2\beta_2}(T_c)_{A_2B_2} (\epsilon c)^\dagger_{f_2 \beta_2 A_2}\zf$
\end{enumerate}

\item $\mathbf{(8,\frac{1}{2},27\oplus 20)}:$

\begin{enumerate}
\item $(T_F)_{fgh}(\EE_A T_a)_{BC}\epsilon_{\alpha\beta} b^\dagger_{f \alpha A} b^\dagger_{g \beta B}b^\dagger_{h \gamma C} b^\dagger_{f_2 \alpha_2 A_2}(T_b)_{A_2B_2} (\epsilon c)^\dagger_{f_2 \alpha_2 B_2}|0_F\rangle $

\item $(T_F)_{fgh}(\EE_A T_a)_{BC}(\tau_i\epsilon)_{\alpha\beta} b^\dagger_{f \alpha A} b^\dagger_{g \beta B}b^\dagger_{h \gamma C} b^\dagger_{f_2 \alpha_2 A_2} (\tau_i)_{\alpha_2\beta_2} (T_b)_{A_1B_2}(\epsilon c)^\dagger_{f_2 \beta_2 B_2}\zf$
\end{enumerate} 

\item $\mathbf{(8,1\otimes\frac{1}{2},27\oplus 20)}:$

\begin{enumerate}
\item $(T_F)_{fgh}(\EE_A T_a)_{BC}(\tau_i\epsilon)_{\alpha\beta} b^\dagger_{f \alpha A} b^\dagger_{g \beta B}b^\dagger_{h \gamma C}b^\dagger_{f_2 \alpha_2 A_2}(T_b)_{A_2B_2} (\epsilon c)^\dagger_{f_2 \alpha_2 B_2}|0_F\rangle $

\item $(T_F)_{fgh}(\EE_A T_a)_{BC}\epsilon_{\alpha\beta} b^\dagger_{f \alpha A} b^\dagger_{g \beta B}b^\dagger_{h \gamma C}b^\dagger_{f_2 \alpha_2 A_2}(\tau_i)_{\alpha_2\beta_2}(T_b)_{A_2B_2} (\epsilon c)^\dagger_{f_2 \beta_2 B_2}|0_F\rangle $

\item $\epsilon_{ijk} (T_F)_{fgh}(\EE_A T_a)_{BC}(\tau_j\epsilon)_{\alpha\beta} b^\dagger_{f \alpha A} b^\dagger_{g \beta B}b^\dagger_{h \gamma C} b^\dagger_{f_2 \alpha_2 A_2} (\tau_k)_{\alpha_2\beta_2} (T_b)_{A_1B_2}(\epsilon c)^\dagger_{f_2 \beta_2 B_2}\zf$
\end{enumerate} 

\item $\mathbf{(8,2\otimes\frac{1}{2},27\oplus 20)}:$

\begin{enumerate}

\item $(T_F)_{fgh}(\EE_A T_a)_{BC}(\tau_i\epsilon)_{\alpha\beta} b^\dagger_{f \alpha A} b^\dagger_{g \beta B}b^\dagger_{h \gamma C} b^\dagger_{f_2 \alpha_2 A_2} (\tau_j)_{\alpha_2\beta_2} (T_b)_{A_1B_2}(\epsilon c)^\dagger_{f_2 \beta_2 B_2}\zf$
\end{enumerate}

\item $\mathbf{(10,\frac{1}{2},1)}:$

\begin{enumerate}

\item $\Delta^\rho_{fgh}(\EE_A)_{BC}(\tau_i\epsilon)_{\alpha\beta}b^\dagger_{f \alpha A} b^\dagger_{g \beta B}b^\dagger_{h \gamma C} b^\dagger_{f_2 \alpha_2 A_2} (\tau_i)_{\alpha_2\beta_2} (\epsilon c)^\dagger_{f_2 \beta_2 A_2}\zf$

\item $\Delta^\rho_{fgh}(\EE_A T_a)_{BC}\epsilon_{\alpha\beta}b^\dagger_{f \alpha A} b^\dagger_{g \beta B}b^\dagger_{h \gamma C} b^\dagger_{f_2 \alpha_2 A_2} (T_a)_{A_2 B_2} (\epsilon c)^\dagger_{f_2 \alpha_2 A_2}\zf$

\item $\Delta^\rho_{fgh}(\EE_A T_a)_{BC}(\tau_i\epsilon)_{\alpha\beta}b^\dagger_{f \alpha A} b^\dagger_{g \beta B}b^\dagger_{h \gamma C} b^\dagger_{f_2 \alpha_2 A_2} (\tau_i)_{\alpha_2\beta_2} (T_a)_{A_2B_2} (\epsilon c)^\dagger_{f_2 \beta_2 B_2}\zf$
\end{enumerate} 

\item $\mathbf{(10,1\otimes\frac{1}{2},1)}:$
\begin{enumerate}
\item $ \Delta^\rho_{fgh}(\EE_A)_{BC}(\tau_i\epsilon)_{\alpha\beta} b^\dagger_{f \alpha A} b^\dagger_{g \beta B}b^\dagger_{h \gamma C}b^\dagger_{f_2 \alpha_2 A_2} (\epsilon c)^\dagger_{f_2 \alpha_2 A_2}|0_F\rangle $

\item $\Delta^\rho_{fgh}(\EE_A T_a)_{BC}\epsilon_{\alpha\beta}b^\dagger_{f \alpha A} b^\dagger_{g \beta B}b^\dagger_{h \gamma C} b^\dagger_{f_2 \alpha_2 A_2} (\tau_i)_{\alpha_2\beta_2} (T_a)_{A_2B_2} (\epsilon c)^\dagger_{f_2 \beta_2 B_2}\zf$

\item $\Delta^\rho_{fgh}(\EE_A T_a)_{BC}(\tau_i\epsilon)_{\alpha\beta}b^\dagger_{f \alpha A} b^\dagger_{g \beta B}b^\dagger_{h \gamma C} b^\dagger_{f_2 \alpha_2 A_2} (T_a)_{A_2B_2} (\epsilon c)^\dagger_{f_2 \alpha_2 B_2}\zf$

\item $\epsilon_{ijk} \Delta^\rho_{fgh}(\EE_A)_{BC}(\tau_j\epsilon)_{\alpha\beta} b^\dagger_{f \alpha A} b^\dagger_{g \beta B}b^\dagger_{h \gamma C}b^\dagger_{f_2 \alpha_2 A_2}(\tau_k)_{\alpha_2\beta_2} (\epsilon c)^\dagger_{f_2 \beta_2 A_2}|0_F\rangle $

\item $\epsilon_{ijk} \Delta^\rho_{fgh}(\EE_A T_a)_{BC}(\tau_j\epsilon)_{\alpha\beta} b^\dagger_{f \alpha A} b^\dagger_{g \beta B}b^\dagger_{h \gamma C}b^\dagger_{f_2 \alpha_2 A_2}(\tau_k)_{\alpha_2\beta_2} (T_a)_{A_2 B_2} (\epsilon c)^\dagger_{f_2 \beta_2 B_2}|0_F\rangle $


\end{enumerate} 

\item $\mathbf{(10,2\otimes\frac{1}{2},1)}:$
\begin{enumerate}
\item $\Delta^\rho_{fgh}(\EE_A)_{BC}(\tau_i\epsilon)_{\alpha\beta} b^\dagger_{f \alpha A} b^\dagger_{g \beta B}b^\dagger_{h \gamma C} b^\dagger_{f_2 \alpha_2 A_2} (\tau_j)_{\alpha_2\beta_2}(\epsilon c)^\dagger_{f_2 \beta_2 A_2}\zf$

\item $\Delta^\rho_{fgh}(\EE_A T_a)_{BC}(\tau_i\epsilon)_{\alpha\beta} b^\dagger_{f \alpha A} b^\dagger_{g \beta B}b^\dagger_{h \gamma C} b^\dagger_{f_2 \alpha_2 A_2}(\tau_j)_{\alpha_2\beta_2}(T_a)_{A_2B_2} (\epsilon c)^\dagger_{f_2 \beta_2 A_2}\zf$
\end{enumerate}

\item $\mathbf{(10,\frac{1}{2},8)}:$
\begin{enumerate}
\item $ \Delta^\rho_{fgh}(\EE_A T_a)_{BC}\epsilon_{\alpha\beta} b^\dagger_{f \alpha A} b^\dagger_{g \beta B}b^\dagger_{h \gamma C}b^\dagger_{f_2 \alpha_2 A_2} (\epsilon c)^\dagger_{f_2 \alpha_2 A_2}|0_F\rangle $
\item $ \Delta^\rho_{fgh}(\EE_A)_{BC}(\tau_i\epsilon)_{\alpha\beta} b^\dagger_{f \alpha A} b^\dagger_{g \beta B}b^\dagger_{h \gamma C}b^\dagger_{f_2 \alpha_2 A_2}(\tau_i)_{\alpha_2\beta_2}(T_a)_{A_2B_2} (\epsilon c)^\dagger_{f_2 \beta_2 B_2}|0_F\rangle $
\item $\Delta^\rho_{fgh}(\EE_A T_a)_{BC}(\tau_i\epsilon)_{\alpha\beta} b^\dagger_{f \alpha A} b^\dagger_{g \beta B}b^\dagger_{h \gamma C}b^\dagger_{f_2 \alpha_2 A_2}(\tau_i)_{\alpha_2\beta_2} (\epsilon c)^\dagger_{f_2 \beta_2 A_2}|0_F\rangle $

\item $f_{abc} \Delta^\rho_{fgh}(\EE_A T_b)_{BC}\epsilon_{\alpha\beta} b^\dagger_{f \alpha A} b^\dagger_{g \beta B}b^\dagger_{h \gamma C}b^\dagger_{f_2 \alpha_2 A_2}(T_c)_{A_2B_2} (\epsilon c)^\dagger_{f_2 \alpha_2 B_2}|0_F\rangle $
\item $d_{abc} \Delta^\rho_{fgh}(\EE_A T_b)_{BC}\epsilon_{\alpha\beta} b^\dagger_{f \alpha A} b^\dagger_{g \beta B}b^\dagger_{h \gamma C}b^\dagger_{f_2 \alpha_2 A_2}(T_c)_{A_2B_2} (\epsilon c)^\dagger_{f_2 \alpha_2 B_2}|0_F\rangle $

\item $f_{abc} \Delta^\rho_{fgh}(\EE_A T_b)_{BC}(\tau_i\epsilon)_{\alpha\beta} b^\dagger_{f \alpha A} b^\dagger_{g \beta B}b^\dagger_{h \gamma C}b^\dagger_{f_2 \alpha_2 A_2}(\tau_i)_{\alpha_2\beta_2}(T_c)_{A_2B_2} (\epsilon c)^\dagger_{f_2 \beta_2 B_2}|0_F\rangle $
\item $d_{abc} \Delta^\rho_{fgh}(\EE_A T_b)_{BC}(\tau_i\epsilon)_{\alpha\beta} b^\dagger_{f \alpha A} b^\dagger_{g \beta B}b^\dagger_{h \gamma C}b^\dagger_{f_2 \alpha_2 A_2}(\tau_i)_{\alpha_2\beta_2}(T_c)_{A_2B_2} (\epsilon c)^\dagger_{f_2 \beta_2 B_2}|0_F\rangle $

\end{enumerate}

\item $\mathbf{(10,1\otimes\frac{1}{2},8)}:$

\begin{enumerate}
\item $\Delta^\rho_{fgh}(\EE_A T_a)_{BC}(\tau_i\epsilon)_{\alpha\beta} b^\dagger_{f \alpha A} b^\dagger_{g \beta B}b^\dagger_{h \gamma C} b^\dagger_{f_2 \alpha_2 A_2} (\epsilon c)^\dagger_{f_2 \alpha_2 A_2}\zf $

\item $\Delta^\rho_{fgh}(\EE_A)_{BC}(\tau_i\epsilon)_{\alpha\beta} b^\dagger_{f \alpha A} b^\dagger_{g \beta B}b^\dagger_{h \gamma C} b^\dagger_{f_2 \alpha_2 A_2}(T_a)_{A_2B_2} (\epsilon c)^\dagger_{f_2 \alpha_2 B_2}\zf $

\item $\Delta^\rho_{fgh}(\EE_A T_a)_{BC}\epsilon_{\alpha\beta} b^\dagger_{f \alpha A} b^\dagger_{g \beta B}b^\dagger_{h \gamma C} b^\dagger_{f_2 \alpha_2 A_2}(\tau_i)_{\alpha_2\beta_2} (\epsilon c)^\dagger_{f_2 \beta_2 A_2}\zf $


\item $f_{abc}\Delta^\rho_{fgh}(\EE_A T_b)_{BC}(\tau_i\epsilon)_{\alpha\beta} b^\dagger_{f \alpha A} b^\dagger_{g \beta B}b^\dagger_{h \gamma C} b^\dagger_{f_2 \alpha_2 A_2} (T_c)_{A_2B_2} (\epsilon c)^\dagger_{f_2 \alpha_2 A_2}\zf $

\item $d_{abc}\Delta^\rho_{fgh}(\EE_A T_b)_{BC}(\tau_i\epsilon)_{\alpha\beta} b^\dagger_{f \alpha A} b^\dagger_{g \beta B}b^\dagger_{h \gamma C} b^\dagger_{f_2 \alpha_2 A_2} (T_c)_{A_2B_2} (\epsilon c)^\dagger_{f_2 \alpha_2 A_2}\zf $

\item $f_{abc}\Delta^\rho_{fgh}(\EE_A T_b)_{BC}\epsilon_{\alpha\beta} b^\dagger_{f \alpha A} b^\dagger_{g \beta B}b^\dagger_{h \gamma C} b^\dagger_{f_2 \alpha_2 A_2}(\tau_i)_{\alpha_2\beta_2} (T_c)_{A_2B_2} (\epsilon c)^\dagger_{f_2 \beta_2 A_2}\zf $

\item $d_{abc}\Delta^\rho_{fgh}(\EE_A T_b)_{BC}\epsilon_{\alpha\beta} b^\dagger_{f \alpha A} b^\dagger_{g \beta B}b^\dagger_{h \gamma C} b^\dagger_{f_2 \alpha_2 A_2}(\tau_i)_{\alpha_2\beta_2} (T_c)_{A_2B_2} (\epsilon c)^\dagger_{f_2 \beta_2 A_2}\zf $

\item $\epsilon_{ijk} \Delta^\rho_{fgh}(\EE_A)_{BC}(\tau_j\epsilon)_{\alpha\beta} b^\dagger_{f \alpha A} b^\dagger_{g \beta B}b^\dagger_{h \gamma C} b^\dagger_{f_2 \alpha_2 A_2} (\tau_k)_{\alpha_2\beta_2} (\epsilon c)^\dagger_{f_2 \beta_2 A_2}\zf $

\item $\epsilon_{ijk} \Delta^\rho_{fgh}(\EE_A)_{BC}(\tau_j\epsilon)_{\alpha\beta} b^\dagger_{f \alpha A} b^\dagger_{g \beta B}b^\dagger_{h \gamma C} b^\dagger_{f_2 \alpha_2 A_2} (\tau_k)_{\alpha_2\beta_2} (T_a)_{A_2B_2}(\epsilon c)^\dagger_{f_2 \beta_2 B_2}\zf $

\item $\epsilon_{ijk} f_{abc} \Delta^\rho_{fgh}(\EE_A T_b)_{BC}(\tau_j\epsilon)_{\alpha\beta} b^\dagger_{f \alpha A} b^\dagger_{g \beta B}b^\dagger_{h \gamma C} b^\dagger_{f_2 \alpha_2 A_2} (\tau_k)_{\alpha_2\beta_2} (T_c)_{A_2B_2} (\epsilon c)^\dagger_{f_2 \beta_2 B_2}\zf $

\item $\epsilon_{ijk} d_{abc} \Delta^\rho_{fgh}(\EE_A T_b)_{BC}(\tau_j\epsilon)_{\alpha\beta} b^\dagger_{f \alpha A} b^\dagger_{g \beta B}b^\dagger_{h \gamma C} b^\dagger_{f_2 \alpha_2 A_2} (\tau_k)_{\alpha_2\beta_2} (T_c)_{A_2B_2} (\epsilon c)^\dagger_{f_2 \beta_2 B_2}\zf $
\end{enumerate}

\item $\mathbf{(10,2\otimes\frac{1}{2},8)}:$
\begin{enumerate}
\item $\Delta^\rho_{fgh}(\EE_A T_a)_{BC}(\tau_i\epsilon)_{\alpha\beta} b^\dagger_{f \alpha A} b^\dagger_{g \beta B}b^\dagger_{h \gamma C} b^\dagger_{f_2 \alpha_2 A_2} (\tau_j)_{\alpha_2\beta_2}(\epsilon c)^\dagger_{f_2 \beta_2 A_2}\zf$

\item $\Delta^\rho_{fgh}(\EE_A)_{BC}(\tau_i\epsilon)_{\alpha\beta} b^\dagger_{f \alpha A} b^\dagger_{g \beta B}b^\dagger_{h \gamma C} b^\dagger_{f_2 \alpha_2 A_2} (\tau_j)_{\alpha_2\beta_2}(T_a)_{A_2B_2}(\epsilon c)^\dagger_{f_2 \beta_2 B_2}\zf$

\item $f_{abc}\Delta^\rho_{fgh}(\EE_A T_b)_{BC}(\tau_i\epsilon)_{\alpha\beta} b^\dagger_{f \alpha A} b^\dagger_{g \beta B}b^\dagger_{h \gamma C} b^\dagger_{f_2 \alpha_2 A_2}(\tau_j)_{\alpha_2\beta_2}(T_c)_{A_2B_2} (\epsilon c)^\dagger_{f_2 \beta_2 A_2}\zf$

\item $d_{abc}\Delta^\rho_{fgh}(\EE_A T_b)_{BC}(\tau_i\epsilon)_{\alpha\beta} b^\dagger_{f \alpha A} b^\dagger_{g \beta B}b^\dagger_{h \gamma C} b^\dagger_{f_2 \alpha_2 A_2}(\tau_j)_{\alpha_2\beta_2}(T_c)_{A_2B_2} (\epsilon c)^\dagger_{f_2 \beta_2 A_2}\zf$
\end{enumerate}

\item $\mathbf{(10,\frac{1}{2},27\oplus 20)}:$

\begin{enumerate}
\item $\Delta^\rho_{fgh}(\EE_A T_a)_{BC}\epsilon_{\alpha\beta} b^\dagger_{f \alpha A} b^\dagger_{g \beta B}b^\dagger_{h \gamma C} b^\dagger_{f_2 \alpha_2 A_2}(T_b)_{A_2B_2} (\epsilon c)^\dagger_{f_2 \alpha_2 B_2}|0_F\rangle $

\item $\Delta^\rho_{fgh}(\EE_A T_a)_{BC}(\tau_i\epsilon)_{\alpha\beta} b^\dagger_{f \alpha A} b^\dagger_{g \beta B}b^\dagger_{h \gamma C} b^\dagger_{f_2 \alpha_2 A_2} (\tau_i)_{\alpha_2\beta_2} (T_b)_{A_1B_2}(\epsilon c)^\dagger_{f_2 \beta_2 B_2}\zf$
\end{enumerate} 

\item $\mathbf{(10,1\otimes\frac{1}{2},27\oplus 20)}:$

\begin{enumerate}
\item $\Delta^\rho_{fgh}(\EE_A T_a)_{BC}(\tau_i\epsilon)_{\alpha\beta} b^\dagger_{f \alpha A} b^\dagger_{g \beta B}b^\dagger_{h \gamma C}b^\dagger_{f_2 \alpha_2 A_2}(T_b)_{A_2B_2} (\epsilon c)^\dagger_{f_2 \alpha_2 B_2}|0_F\rangle $

\item $\Delta^\rho_{fgh}(\EE_A T_a)_{BC}\epsilon_{\alpha\beta} b^\dagger_{f \alpha A} b^\dagger_{g \beta B}b^\dagger_{h \gamma C}b^\dagger_{f_2 \alpha_2 A_2}(\tau_i)_{\alpha_2\beta_2}(T_b)_{A_2B_2} (\epsilon c)^\dagger_{f_2 \beta_2 B_2}|0_F\rangle $

\item $\epsilon_{ijk} \Delta^\rho_{fgh}(\EE_A T_a)_{BC}(\tau_j\epsilon)_{\alpha\beta} b^\dagger_{f \alpha A} b^\dagger_{g \beta B}b^\dagger_{h \gamma C} b^\dagger_{f_2 \alpha_2 A_2} (\tau_k)_{\alpha_2\beta_2} (T_b)_{A_1B_2}(\epsilon c)^\dagger_{f_2 \beta_2 B_2}\zf$
\end{enumerate} 

\item $\mathbf{(10,2\otimes\frac{1}{2},27\oplus 20)}:$

\begin{enumerate}

\item $\Delta^\rho_{fgh}(\EE_A T_a)_{BC}(\tau_i\epsilon)_{\alpha\beta} b^\dagger_{f \alpha A} b^\dagger_{g \beta B}b^\dagger_{h \gamma C} b^\dagger_{f_2 \alpha_2 A_2} (\tau_j)_{\alpha_2\beta_2} (T_b)_{A_1B_2}(\epsilon c)^\dagger_{f_2 \beta_2 B_2}\zf$
\end{enumerate} 

\end{enumerate}

\subsection{Gluon states}

As gluon trial wavefunctions, we choose eigenstates of a $24$-dimensional harmonic oscillator, defined in (\ref{hosc}).
Introducing oscillator creation and annihilation operators, 
\begin{equation}
A_{ia}=\frac{1}{\sqrt{2}} \left(i\Pi_{ia}+M_{ia}\right),
\end{equation}
\begin{equation}
A^\dagger_{ia}=\frac{1}{\sqrt{2}} \left(-i\Pi_{ia}+M_{ia}\right),
\end{equation}
we create oscillator eigenstates by successive operations of $A^\dagger$ on the oscillator vacuum $\zb$. These states can be organized into real representations of $\text{Ad } SU(3)$, formed by decomposing tensor products of the adjoint representation.

To write gluon states, we define a matrix $B$ whose elements are oscillator creation operators.
\begin{equation}
B_{ia}=A^\dagger_{ia}
\end{equation}
A gluon state with a given number of oscillators can be expressed as a function of $B$ acting on the oscillator vacuum $\zb$. We define $(D_a)_{bc}\equiv d_{abc}$ and $(F_a)_{bc}\equiv f_{abc}$. The gluon states are organized according to their spin and colour. To make varational estimates of spin-$1$ mesons and spin-$\frac{3}{2}$ baryons, we need the gluon states to have spin up to $3$. Again, for notational convenience, a spin-3 state is written with three free adjoint indices $i,j,k$, and the spin-3 state is obtained from a state $|\phi_{ijk}\rangle$ as
\begin{equation}
|\phi^{(3)}_{ijk}\rangle=|\phi_{(ijk)}\rangle-\frac{2}{5}\sum_l \delta_{(ij}[|\phi_{k)ll}\rangle+|\phi_{l k)l}\rangle+|\phi_{ll k)}\rangle]
\end{equation}
where the brackets around the indices $(ijk)$ denotes symmetrization in $i,j,k$.

\subsubsection*{Gluon states}
\begin{enumerate}

\item \textbf{(0,1)}:
\begin{enumerate}
\item $\zb$
\item (Tr $BB^T)\zb$
\item $\epsilon_{ijk}f_{abc}B_{ia}B_{jb}B_{kc}\zb$
\item (Tr $BB^T)^2\zb$
\item (Tr $BB^TBB^T)\zb$
\item (Tr $BD_aB^T$)(Tr $BD_aB^T)\zb$
\end{enumerate}

\item \textbf{(1,1)}:
\begin{enumerate}
\item $B_{ia}$(Tr $BD_aB^T)\zb$
\item $\epsilon_{jkl}(BD_cB^T)_{ij}(BF_cB^T)_{kl}\zb$
\end{enumerate}

\item \textbf{(2,1)}:
\begin{enumerate}
\item $(BB^T)_{ij}\zb$
\item (Tr $BB^T)(BB^T)_{ij}\zb$
\item $(BB^TBB^T)_{ij}\zb$
\item (Tr $BD_aB^T) (BD_aB^T)_{ij}\zb$
\end{enumerate}

\item \textbf{(0,8)}:
\begin{enumerate}
\item (Tr $BD_a B^T)\zb$
\item $\epsilon_{ijk}(BF_cB^T)_{ij}(BD_c)_{ka}\zb$
\item (Tr $BB^T)$(Tr $BD_aB^T)\zb$
\item (Tr $B^TBD_aB^TB)\zb$
\item (Tr $BD_bB^T) (B^TB)_{ab}\zb$
\item (Tr $BD_bD_aB^T)$ (Tr $BD_bB^T)\zb$
\end{enumerate}

\item \textbf{(1,8)}:
\begin{enumerate}
\item $B_{ia}\zb$
\item $\epsilon_{ijk}f_{abc}B_{jb}B_{kc}\zb$
\item $B_{ia}$(Tr $BB^T)\zb$
\item $(BB^TB)_{ia}\zb$
\item Tr $(BD_cB^T)(BD_c)_{ia}\zb$
\item Tr $(BD_cB^T)(BF_c)_{ia}\zb$

\item $\epsilon_{jkl}f_{bcd}B_{jb}B_{kc}B_{ld} B_{ia}\zb$
\item $\epsilon_{jkl}(BB^T)_{ij}(BF_aB^T)_{kl}\zb$
\item $\epsilon_{jkl} (BD_aD_bB^T)_{ij}(BF_bB^T)_{kl}\zb$
\item $\epsilon_{jkl} (BF_aD_bB^T)_{ij}(BF_bB^T)_{kl}\zb$
\item $\epsilon_{ijk} (BF_aB^T)_{jk}$(Tr $(BB^T)\zb$
\item $\epsilon_{ijk}$(Tr $BD_bB^T)B_{jc}B_{ka}\zb$

\item $\epsilon_{ijk}$(Tr $BD_bB^T)(BF_bD_aB^T)_{jk}\zb$
\item $\epsilon_{ijk}$(Tr $BD_bB^T)(BF_bF_aB^T)_{jk}\zb$
\end{enumerate}

\item \textbf{(2,8)}:
\begin{enumerate}
\item $(BD_aB^T)_{ij}\zb$
\item $\epsilon_{ilm}(BB^T)_{jl}B_{ma}\zb$
\item $\epsilon_{ilm}(BF_cB^T)_{jl}(BD_c)_{ma}\zb$
\item $\epsilon_{ilm}(BF_cB^T)_{jl}(BF_c)_{ma}\zb$

\item (Tr $BD_bB^T)B_{ia}B_{jb}\zb$
\item $(BB^T)_{ij}$(Tr $BD_aB^T)\zb$
\item (Tr $BB^T)(BD_aB^T)_{ij}\zb$
\item $(BB^TBD_aB^T)_{ij}\zb$
\item $(BB^TBF_aB^T)_{ij}\zb$
\item (Tr $BF_bF_aB^T)(BD_bB^T)_{ij}\zb$
\item (Tr $BD_bD_aB^T)(BF_bB^T)_{ij}\zb$
\item (Tr $BD_bF_aB^T)(BF_bB^T)_{ij}\zb$
\item (Tr $BD_bB^T)(BD_bD_aB^T)_{ij}\zb$
\item (Tr $BD_bB^T)(BD_bF_aB^T)_{ij}\zb$
\item (Tr $BD_bB^T)(BF_bD_aB^T)_{ij}\zb$
\item (Tr $BD_bB^T)(BF_bF_aB^T)_{ij}\zb$
\end{enumerate}

\item \textbf{(0,27$\oplus$20)}:
\begin{enumerate}
\item $(B^TB)_{ab}\zb$
\item$\epsilon_{ijk}(BF_aB^T)_{ij}B_{kb}\zb$
\item (Tr $B^TB)(B^TB)_{ab}\zb$
\item $(B^TBB^TB)_{ab}\zb$
\item (Tr $B^TBD_c) (B^TBD_c)_{ab}\zb$
\item (Tr $B^TBD_c) (B^TBF_c)_{ab}\zb$
\end{enumerate}

\item \textbf{(1,27$\oplus$20)}:
\begin{enumerate}
\item $\epsilon_{ijk}B_{ja}B_{kb}\zb$
\item (Tr $BD_aB^T)B_{ib}\zb$
\item $(BF_aB^TB)_{ib}\zb$
\item $(BD_aB^TB)_{ib}\zb$

\item $\epsilon_{i_1j_1k_1}(BF_cB^T)_{i_1j_1}(BD_c)_{k_1a}B_{ib}\zb$
\item $\epsilon_{ijk}$(Tr $BB^T)B_{ja}B_{kb}\zb$
\item $\epsilon_{ijk}B_{ja}(BB^TB)_{kb}\zb$
\item $\epsilon_{ijk}$(Tr $BD_cB^T)(BD_c)_{ja}B_{kb}\zb$
\item $\epsilon_{ijk}(BD_c)_{ja}(BD_cB^TB)_{kb}\zb$
\item $\epsilon_{ijk}$(Tr $BD_cB^T)(BF_c)_{ja}B_{kb}\zb$
\item $\epsilon_{ijk}(BF_cB^T)_{jk}(B^TBD_c)_{ab}\zb$
\end{enumerate}

\item \textbf{(2,27$\oplus$20)}:
\begin{enumerate}
\item $B_{ia}B_{jb}\zb$

\item $\epsilon_{ilm}(BF_aB^T)_{lm}B_{jb}\zb$
\item $\epsilon_{ilm}(BD_aB^T)_{lm}B_{jb}\zb$

\item (Tr $BB^T)B_{ia}B_{jb}\zb$
\item $(BB^T)_{ij}(B^TB)_{ab}\zb$
\item $(BB^TB)_{ia}B_{jb}\zb$
\item (Tr $BD_cB^T)(BD_c)_{ia}B_{jb}\zb$
\item $(BD_cB^T)_{ij}(B^TBD_c)_{ab}\zb$
\item (Tr $BD_cB^T)(BF_c)_{ia}B_{jb}\zb$
\item $(BD_cB^TB)_{ia}(BF_c)_{jb}\zb$
\item $(BF_cB^T)_{ij}(BD_cB^T)_{ab}\zb$

\end{enumerate}

\item \textbf{(3,1)}:
\begin{enumerate}
\item $d_{abc}B_{ia}B_{jb}B_{kc}\zb$
\end{enumerate}

\item \textbf{(3,8)}:
\begin{enumerate}
\item $(BB^T)_{ij}B_{ka}\zb$
\item $\epsilon_{ilm}(BB^T)_{jl}(BD_aB^T)_{km}\zb$
\item $\epsilon_{ilm}(BB^T)_{jk}(BF_aB^T)_{jm}\zb$
\item $\epsilon_{ilm}(BD_bB^T)_{jk}(BD_bD_aB^T)_{lm}\zb$
\item $\epsilon_{ilm}(BD_bB^T)_{jk}(BD_bF_aB^T)_{jm}\zb$
\end{enumerate}

\item \textbf{(3,27$\oplus$ 20)}:
\begin{enumerate}
\item $(BD_a B^T)_{ij}B_{kc}\zb$
\item $\epsilon_{ilm}(BB^T)_{jl}B_{ka}B_{mb}\zb$
\item $\epsilon_{ilm}(BD_bB^T)_{jl}(BD_b)_{ka}B_{mb}\zb$
\item $\epsilon_{ilm}(BD_bB^T)_{jl}(BF_b)_{ka}B_{mb}\zb$
\end{enumerate}

\end{enumerate}

\end{document}